\documentclass[fleqn,usenatbib]{mnras}
\usepackage{fix-cm}
\usepackage{newtxtext,newtxmath}

\usepackage[T1]{fontenc}

\DeclareRobustCommand{\VAN}[3]{#2}
\let\VANthebibliography\thebibliography
\def\thebibliography{\DeclareRobustCommand{\VAN}[3]{##3}\VANthebibliography}

\usepackage{graphicx}	
\usepackage{amsmath}	
\usepackage[dvipsnames]{xcolor} 
\usepackage[normalem]{ulem}
\usepackage{cancel}
\usepackage{placeins}
\newcommand{\ignore}[1]{}

\usepackage{graphicx}	
\usepackage{amsmath}	
\usepackage{upgreek}
\usepackage{booktabs} 
\usepackage{siunitx} 
\usepackage{bm}    
\title[Ionized outflows in CGCG\:012-070]{Ionized Gas Outflows and Shock-Heated Emission in the Highly Inclined Active Galaxy CGCG 012-070}
\author[Ramos Vieira et al.]{
Lucas Ramos Vieira,$^{1,2}$\thanks{E-mail: lucas.vieira@ifc.edu.br}
Rogemar A. Riffel,$^{1,3}$\thanks{E-mail: rogemar@ufsm.br}
Rogério Riffel,$^{4}$, Oli L. Dors,$^{5}$ Marina Bianchin,$^{6}$ 
\newauthor
Thaisa Storchi-Bergmann$^{4}$
\\
$^{1}$Departamento de F\'isica, CCNE, Universidade Federal de Santa Maria, 97105-900 Santa Maria, RS, Brazil
\\
$^{2}$ Instituto Federal Catarinense - Campus Concórdia, 89703-720, Concórdia, SC, Brazil\\
$^{3}$Centro de Astrobiología (CAB), CSIC-INTA, Ctra. de Ajalvir km 4, Torrejón de Ardoz, E-28850, Madrid, Spain\\
$^{4}$Departamento de Astronomia, IF, Universidade Federal do Rio Grande do Sul, CP 15051, 91501-970, Porto Alegre, RS, Brazil\\ 
$^{5}$Universidade do Vale do Para\'iba, Av. Shishima Hifumi, 2911, 12244-000, S\~ao Jos\'e dos Campos, SP, Brazil \\
$^{6}$Department of Physics and Astronomy, 4129 Frederick Reines Hall, University of California, Irvine, CA 92697, USA\\
}

\date{Accepted XXX. Received YYY; in original form ZZZ}

\pubyear{\the\year{}}

\begin{document}
\label{firstpage}
\pagerange{\pageref{firstpage}--\pageref{lastpage}}
\maketitle

\begin{abstract}
\noindent 
Active Galactic Nuclei (AGNs) exhibit excess mid-infrared H$_2$ emission compared to star-forming galaxies, likely driven by outflows and shocks inferred from integrated spectra. We present optical IFU observations of the central 2\,kpc of the AGN host CGCG\,012-070, selected for its pronounced H$_2$ emission excess, to map stellar and gas kinematics. The stellar velocity field is well described by a rotating disc with a line of nodes at $103^\circ \pm 4^\circ$, with the northwest side approaching and the southeast side receding. Gas kinematics, traced by strong emission lines, show two components: a narrow one ($\sigma \lesssim 200\,{\rm km\,s^{-1}}$) in the disc plane following stellar motions, and a broad ($\sigma \gtrsim 300\,{\rm km\,s^{-1}}$) associated with outflows within the inner $\sim$1\,kpc. Disc gas emission is mainly driven by AGN photoionization, while the outflow also includes shock-heated gas, as indicated by flux ratio diagnostics. The outflows are radiatively driven, with a mass-outflow rate of $(0.067 \pm 0.026)\,M_{\odot}\,{\rm yr^{-1}}$ and a kinetic coupling efficiency of 0.07\%, potentially redistributing gas and contributing to maintenance-mode feedback in CGCG\,012-070. Our results provide further evidence that the warm H$_2$ emission excess in nearby AGN is associated with shocks produced by outflows. Observations of other gas phases, such as cold molecular gas, are necessary to gain a more comprehensive understanding of the impact of the outflows on the host galaxy.
\end{abstract}


\section{Introduction}
The majority of massive galaxies host a supermassive black hole (SMBH) at their core.  At certain periods in the evolution of a galaxy, an adequate supply of gas may gather and form an accretion disc around the SMBH, emitting largely non-thermal radiation \citep{kormendy13}. This phase, characterized by intense radiation and energetic processes, is referred to as an Active Galactic Nucleus (AGN). Semi-analytical models and hydrodynamical simulations have incorporated AGNs as a key mechanism for modulating supermassive black hole growth and regulating star formation (SF) in massive galaxies \citep{ Benson_2003,  Somerville_2008, Madau_2014, Harrison2024}. Such a process is essential to successfully reproduce the observed local mass functions and the colour bimodality of galaxies \citep{Harrison2024}. 

The impact of an AGN on its host galaxy spans multiple scales. Both black hole accretion (AGN activity) and star formation depend on the availability of gas. Therefore, a key challenge is to understand the relation between inflows of gas that fuel the AGN (feeding processes) and the feedback mechanisms that regulate it \citep{Harrison2024, Storchi2019}. Feedback processes may consist of the release of jets from the inner edge of the accretion disc, winds originating from its outer regions and radiation emitted by the hot gas in the disc or its corona, all of which can produce a significant change in the host galaxy as a result of nuclear activity \citep{Elvis2000, king2003}. Analytical models suggest that a fraction of the luminosity produced by powerful AGN can be utilized to generate winds, establishing scaling relations between SMBH mass and their host galaxy properties \citep[e.g.][]{Silk1998, king2003, King_2015, Fiore2017}. Nonetheless, the impact of AGN feedback on these scaling relations remains a subject of discussion \citep{Peng2007, Graham2023, Riffel2024}. Furthermore, the fuel for AGN feeding may have a significant amount of stellar recycled gas that is released in the nuclear region of these galaxies \citep{Choi_2024, Riffel_Rogerio_2024}.

Integral field spectroscopy (IFS) is an ideal tool for exploring the physical properties of stellar and gas components in the central regions of galaxies \citep[e.g.][]{sarzi_2010, Sanchez_2020, Sanchez_2021, ruschel-dutra21, Ricci2023, Riffel2023}, as it enables spatially resolved studies of gas and stellar properties, providing a detailed reconstruction of the star formation history of galaxies and AGN properties. Through IFS it is possible to investigate how various factors influence star formation, from AGN activity to galaxy interactions and environmental effects \citep{Sharp_2010, Goddard2016}. 

\citet{Riffel2020} used mid-infrared and optical data to identify galaxies that can be hosts of strong molecular gas outflows. The sample was obtained by cross-matching 2015 galaxies  with Spitzer mid-infrared spectra from \citet{lambrides19} and SDSS spectroscopy \citep{Gunn_2006, Blanton_2017}, resulting in 309 galaxies with both datasets. Among them, 115 show emission of $\mathrm{H_2\,S(3)~9.665 \micron}$ and PAH~$11.3 \micron$. Spitzer observations reveal that AGN hosts can exhibit enhanced $\text{H}_2$ emission compared to star-forming regions \citep{lambrides19, Petric18, hill14}. This excess is likely due to shocks from AGN-driven outflows, supported by the correlation between the $\mathrm{H_2\,S(3)/PAH\,11.3 \micron}$ ratio and shock indicators such as log${\rm ([\mathrm{O\,I}]\,\uplambda6300/H\alpha)}$ and the $[\mathrm{O\,I}]$ velocity dispersion \citep{Riffel2020}. From single-aperture studies, CGCG\,012-070, NGC\,3884, and UGC\,8782 are the AGN hosts at $z < 0.05$ with the highest [O\,{\sc i}] velocity dispersion and $\mathrm{H_2\,S(3)/PAH\,11.3 \micron}$ ratio. To investigate the origin of the $\mathrm{H}_2$ emission excess and the evidence of shocks, \citet{Riffel2025} used JWST/MIRI medium-resolution spectroscopy for the three galaxies. They concluded that the excess is associated with shock heating, produced either by warm molecular outflows or by radio jet–ISM interactions \citep{Costa-Souza_2024,Riffel2025}.

In addition to studies on the warm molecular gas phase using JWST data for three galaxies mentioned above, the hotter ionized gas counterpart is being analysed using optical integral field unit (IFU) data from the Gemini Multi-Object Spectrograph (GMOS). So far, results on the ionized gas emission structure and kinematics have been published for two galaxies, UGC\:8782 \citep{Riffel2023} and NGC\:3884 \citep{Riffel2024}, which mapped the structure and kinematics of both stellar and gas emission. For UGC\,8782, located at $z = 0.045$, the stars present ordered rotation following the orientation of the large-scale disc of the galaxy, while the gas shows disturbed kinematics with a narrow emission-line component $(\upsigma \lesssim 200\ \mathrm{km\ s^{-1}})$ associated with the gas in the disc and a broad component $(\upsigma \gtrsim 200\ \mathrm{km\ s^{-1}})$ produced by ionized gas outflows. Optical emission-line ratio diagnostic diagrams indicate that the ionized gas emission within the inner 2 kpc of this galaxy is primarily attributed to AGN photoionization. In contrast, the outflow emission exhibits an additional contribution from shock excitation, caused by the interaction between the radio jet and the surrounding gas. The broad component is consistently blueshifted by approximately 150–500 $\mathrm{km\ s^{-1}}$ relative to the systemic velocity of the galaxy across all observed regions. The kinetic coupling efficiency of the ionized outflows ranges from 1 to 3 percent, suggesting that they may be sufficiently strong to affect star formation in the host galaxy \citep[e.g.][]{hopkins_elvis10}. 

Regarding NGC\,3884, a Sab spiral galaxy \citep{Tully_2000} with a redshift of $z = 0.023$ (e.g. \citealt{Rines_2016}), its nuclear activity has been classified as a LINER with a faint H$\alpha$ emission from the broad line region (BLR; \citealt{veron06}). \citet{Riffel2024} reported the following findings for NGC 3884: stellar kinematics dominated by a rotating disc component; emission-line ratio diagnostic diagrams show that AGN photoionization is the main gas excitation mechanism; gas kinematics reveal two components: a narrow emission-line from the gas emission in the disc and a broad one from AGN-driven ionized outflows, with a kinetic power of $\sim0.06\% $ of the AGN bolometric.  
Their results suggest that the outflow is strong enough to suppress over $10\%$ of the star formation of the host galaxy \citep{Almeida2023}. 

As an extension of the studies presented in \citet{Riffel2023} and \citet{Riffel2024} for the two nearby galaxies NGC\,3884 and UGC\,8782, in this work we use optical IFS observations to spatially resolve the emission structure and kinematics of the ionized and neutral gas in the central region of the galaxy CGCG\,012-070. It is an Sb spiral galaxy that shows an inclination nearly edge-on, with axial ratio $b/a \approx 0.25$, located at a redshift $z = 0.0481$ \citep{Kautsch2006}. Its nuclear activity is classified as Seyfert 2 \citep{veron06, Kautsch2006}. CGCG\,012-070 exhibits faint radio emission, with a 1.4 GHz power of $P_{1.4} \approx 2.4 \times 10^{22}$ W Hz$^{-1}$, as measured by the Faint Images of the Radio Sky at Twenty centimeters (FIRST) survey \citep{Lofthouse2018, Becker1995}.

This paper is structured as follows: Section~\ref{sec: observations} provides details on the observations, data reduction procedures, and analysis methods. Section~\ref{sec:results} presents the main results, which are discussed in Section~\ref{sec: discussion}. Finally, Section~\ref{sec: conclusions} summarizes our key conclusions.

\section{Observations and Measurements}
\label{sec: observations}

We use IFU data obtained with the Gemini-South GMOS instrument \citep{hook04} using the B600 grating, to map the gas and stellar properties in the central region of CGCG\,012-070.  The observations were performed on 2022 January 05 under the program code GS-2021B-Q-220 (PI: Riffel, R. A.) and consisted of four exposures of 1260 sec each, two of them centred at 615 nm and two at 590 nm. We used GMOS-IFU in the one-slit mode, resulting in a $\mathrm{3.5 \times 5.0 \, arcsec^2}$ field of view and covering an observed spectral range of 4450-7400\,\AA.   During the observations, the major axis of the IFU was oriented along the position angle PA=105$^\circ$, approximately along the major axis of the galaxy.

The data reduction was performed with the {\sc gemini.gmos iraf} package. The procedure included the subtraction of the bias level, trimming and flat-field correction, background subtraction for each science data frame, quantum efficiency correction, sky subtraction, and wavelength calibration. We applied the {\sc lacos} code \citep{van_Dokkum_2001} to further remove cosmic ray contamination, and finally we performed the flux calibration using observations of the standard star LTT 9491 to construct the sensitivity function. The final data cube, with an angular sampling of $\mathrm{0.1 \times 0.1 arcsec^2}$, was constructed by median combining the data cubes for each exposure using the peak of the continuum emission as reference to align them. The velocity resolution is $\sim85$ km\:s$^{-1}$, as measured from the full width at half-maximum (FWHM) of a typical emission line in the CuAr spectra used in the wavelength calibration and the seeing is $\sim$0.7 arcsec as estimated from the FWHM of flux distributions of field stars, which corresponds to a spatial resolution of $\sim$715 pc at the distance of CGCG\,012-070. After completion of the data reduction, we identified a telluric absorption line at $\sim$ 6553 \AA, likely due to the association with the oxygen B band \citep{Smette_2015}. We fitted this feature by a single Gaussian and subtracted it from the observed data cube. 

The measurements were performed using the same procedures described in \citet{Riffel2023}. In summary, we use the Penalized Pixel-Fitting (PPXF) method \citep{cappellar04, Cappellari2017, Cappellari2023} to fit the contribution of the underlying stellar population to the observed spectra and obtain measurements of the stellar kinematics, using the MILES-HC library \citep{Westfall_2019} as spectral templates. 

We employ the IFSCUBE package \citep{ruschel-dutra21} to fit the observed emission-line profiles using Gaussian curves. The fitting procedure is carried out after subtracting the stellar component contribution from the observed spectra, allowing the code to include up to two Gaussian components (narrow and broad) per emission line, as indicated by a visual inspection of the spectra. The CUBEFIT routine from the IFSCUBE package is used to simultaneously fit the following emission lines: H$\beta$, [O\,{\sc{iii}}]\, $\uplambda \uplambda$4959,\,5007, [N\,{\sc{i}}]\,$\uplambda$5200, [O\,{\sc{i}}]\,$\uplambda\uplambda$6300,\,6363, [N\,{\sc{ii}}]\,$ \uplambda \uplambda$6548,\,6583, H$\alpha$, and [S\,{\sc{ii}}]\,$\uplambda \uplambda$6717,\,6731.

Initial guesses for the Gaussian amplitude, centroid velocity, and velocity dispersion are based on the fitting of the nuclear emission line profiles. Once the nuclear emission-line profiles are successfully fitted, the routine proceeds to fit the surrounding spaxels following a radial spiral pattern. The initial parameters for these fits are then taken from the best-fitting values obtained for spaxels located within 0.3 arcsec of the spaxel being analyzed, as defined by the \texttt{refit} parameter. If the amplitude of one of the fitted Gaussian components is lower than three times the standard deviation of the adjacent continuum, only a single component is considered.  

To account for any remaining continuum emission, a third-order polynomial is also included in the fitting procedure, applied before fitting the emission lines. During the fit, for each component, the velocity and velocity dispersion ($\upsigma$) are tied within three kinematic groups, organized as follows: H$\beta$, [O\,{\sc{iii}}]\,$\uplambda$4959 and [O\,{\sc{iii}}]\,$\uplambda$5007; [N\,{\sc{i}}]\,$\uplambda$5200;  [O\,{\sc{i}}]\,$\uplambda $6300, [O\,{\sc{i}}]\,$\uplambda$6363, [N\,{\sc{ii}}]\,$ \uplambda$6548, H$\upalpha$, [N\,{\sc{ii}}]\,$ \uplambda$6583, [S\,{\sc{ii}}]\,$ \uplambda$6717 and [S\,{\sc{ii}}]\,$ \uplambda$6731. Additionally, the intensity ratios [O\,{\sc{iii}}]\,$ \uplambda$5007/[O\,{\sc{iii}}]\,$ \uplambda$4959 and [N\,{\sc{ii}}]\,$\uplambda$6583/[N\,{\sc{ii}}]\,$\uplambda$6548 are fixed at their theoretical values of 2.98 and 2.96, respectively \citep{Osterbrock2006, DOJCINOVIC2023}.

\begin{figure*}
    \centering
    \includegraphics[width=0.99\textwidth, trim=3.1cm 0.5cm 4.0cm 1cm, clip]{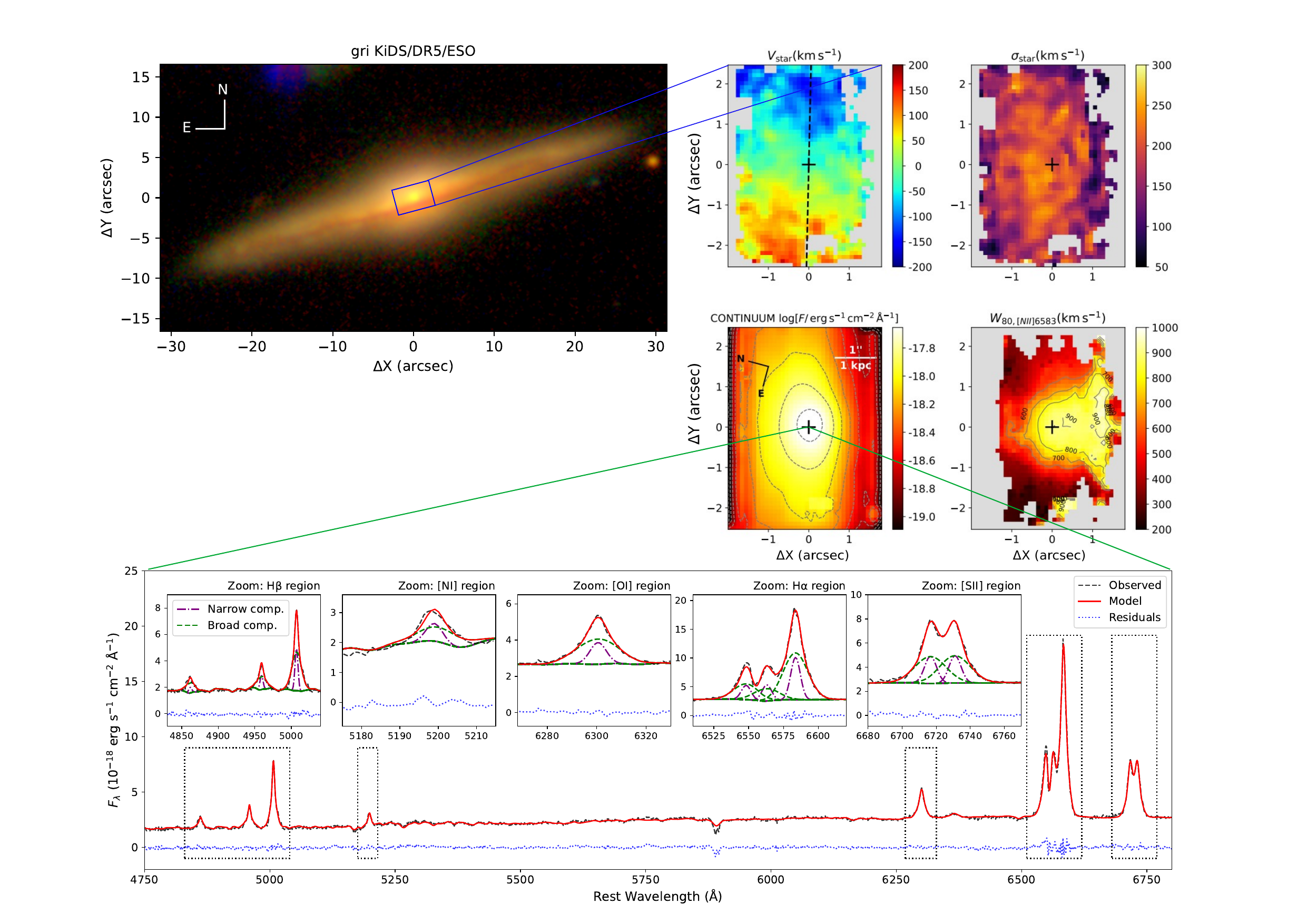} 
    \caption{The top-left panel shows a composite image of the $\sim$ 30 $\times$ 60 $\mathrm{arcsec^2}$ in $g$ (477 nm, shown in blue), $r$ (653 nm, in green), and $i$ (763 nm, in red) bands of CGCG\,012-070 using data from the KiDS/DR5/ESO archive \citep{Wright_2024}. The top-centre and right panels show the stellar velocity and velocity dispersion maps, respectively. Grey regions in these maps correspond to regions with uncertainties in velocity larger than 30 km $\mathrm{s^{-1}}$  or $\upsigma$ larger than 50 km $\mathrm{s^{-1}}$. The systemic velocity of the galaxy has been subtracted from the observed velocity map. The scale bars show the velocities in km $\mathrm{s^{-1}}$ and the black dashed line represent the line of nodes, obtained by fitting the observed velocities by a rotation disc model (Sec.~\ref{sec: disccomponent}). The central panel shows a GMOS continuum image obtained within a spectral window of 100 \AA \, width centred at 6100 \AA. The scale bar shows the continuum flux values in logarithmic units of erg $\mathrm{s^{-1}}$ $\mathrm{cm^{-2}}$ \AA$^{-1}$.  The central panel and the middle-right panel shows the maps for continuum flux and $\mathrm{W_{80}}$ map for the [N\,{\sc{ii}}]\,$ \uplambda$6583 emission line, respectively, with the scale bar indicating the values in km\,$\mathrm{s^{-1}}$. Grey regions indicate areas where the corresponding emission line is not detected with S/N$\geq$3.  The bottom plot shows the observed spectrum for the nuclear spaxel in black, the best-fitting model in red, and the residuals in blue. The strongest emission lines are identified. The five plots show a zoom into the emission line regions to illustrate the multi-Gaussian fit, including the stellar population contribution. The broad line components are shown in dashed green and the narrow components are in dashed-dot purple.}
    \label{fig:large}
\end{figure*}

\section{Results}
\label{sec:results}

\begin{figure*}
    \centering
    \includegraphics[width=0.71\textwidth, trim=0.2cm 0.2cm 0.2cm 1cm, clip]{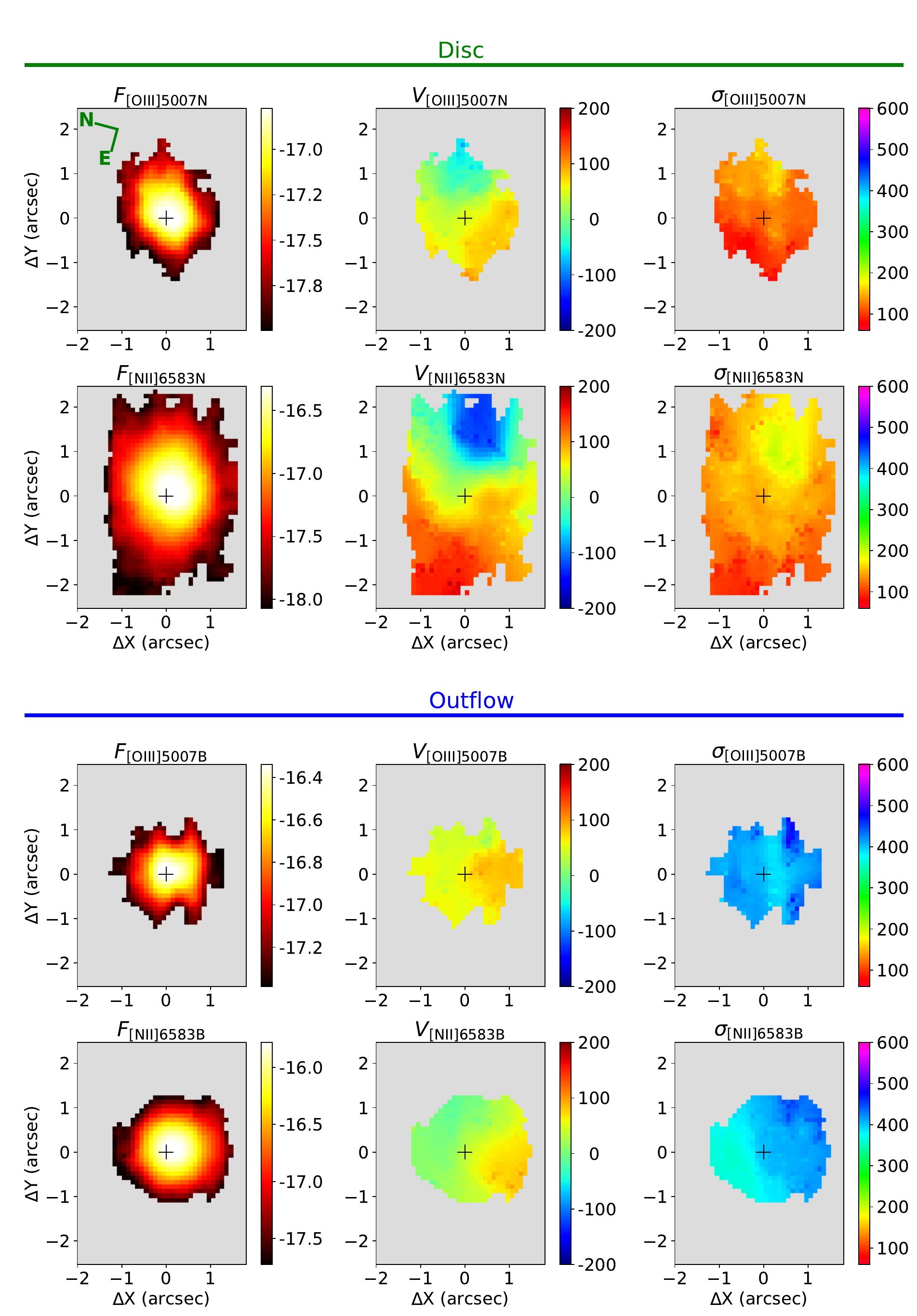}
    \vspace{0.3cm} 
    \caption{The panels display the flux distributions and kinematic maps for the [O\,{\sc{iii}}]\,$\uplambda$5007 and [N\,{\sc{ii}}]\,$\uplambda$6583 emission lines. For each line, the columns, from left to right, show the flux distribution, centroid velocity, and velocity dispersion maps. The top two rows correspond to the narrow component, while the bottom two rows represent the broad component. The scale bars indicate line fluxes in logarithmic units of $\mathrm{erg\,s^{-1}\,cm^{-2}}$, with velocities and velocity dispersions in $\mathrm{km\,s^{-1}}$. Central crosses mark the position of the continuum peak, while the spatial orientation is provided in the top-left panel. Gray regions indicate spaxels where the emission line is not detected above three times the standard deviation of the continuum in spectral regions adjacent to each line. }
    \label{fig:maps1_novo}

\end{figure*}

\begin{figure*}    
    \includegraphics[width=0.99\textwidth]{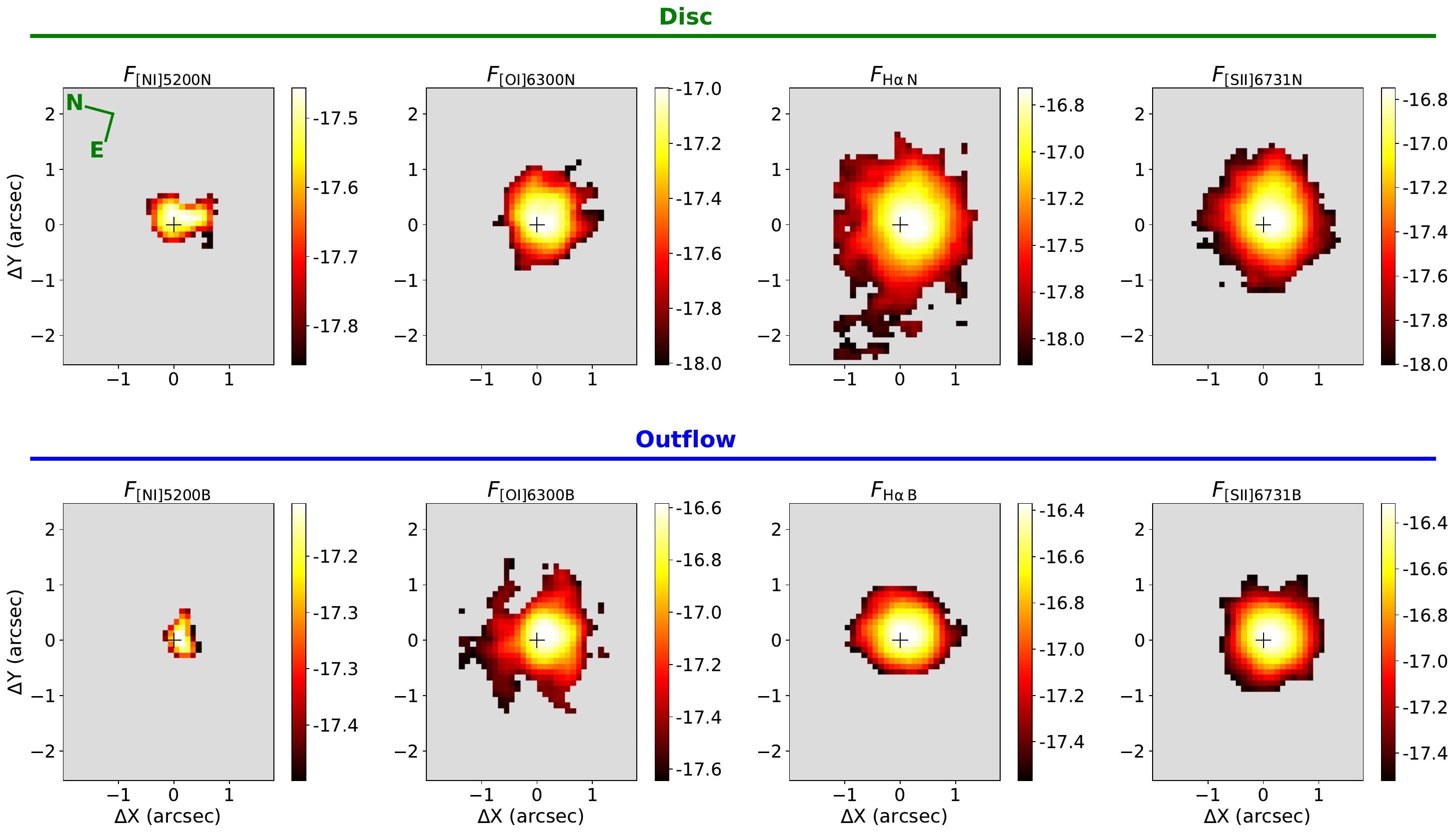}
    
    \caption{From left to right, the panels show the flux distributions for the [N\,{\sc{i}}]\,$\uplambda$5200, [O\,{\sc{i}}]\,$\uplambda$6300, H$\upalpha$, and [S\,{\sc{ii}}]\,$\uplambda$6731 emission lines. The top and bottom panels correspond to the narrow and broad components, respectively. The scale bars represent line fluxes in logarithmic units of $\mathrm{erg\,s^{-1}\,cm^{-2}}$.}
    \label{fig:maps2_novo}
\end{figure*}

Figure~\ref{fig:large} shows a broad overview of the GMOS observations of CGCG-012-070. The top-left panel shows a large-scale composite image constructed using data from the Kilo Degree Survey (KiDS/DR5/ESO) archive \citep{Wright_2024}, showing in blue the g-band (477 nm) image, in green the r-band (653 nm) image, and in red the i-band (763 nm) image. The spatial orientation is shown in the top-left corner of the picture, and the inner blue rectangle shows the GMOS-IFU field of view (FoV). The top-centre and right panels show the stellar velocity and stellar velocity dispersion ($\sigma_\star$) maps, respectively. The stellar velocity field is corrected by the heliocentric systemic velocity of the galaxy, of $V_s=14\,480$ km\,s$^{-1}$, as derived by fitting the observed velocity field by a rotation disc model (Sec. \ref{sec: disccomponent}). 
The stellar velocity field exhibits a disc-like rotation pattern with the west side of the disc approaching and the east side receding, with a projected velocity amplitude of $\sim \mathrm{200 \, km\,s^{-1}}$.
The stellar velocity dispersion map presents values in the range of 70--270 km $\mathrm{s^{-1}}$. 
The middle-centre panel of Fig.~\ref{fig:large} presents the Gemini-GMOS continuum image derived by calculating the mean flux in a 100\,\AA \: window centred at 6100\,\AA. The continuum distribution is elongated approximately  along the east-west direction, consistent with the orientation of the large scale disc \citep{Adelman2008}. The middle-right panel shows the $\mathrm{W_{80}}$ map for the [N\,{\sc{ii}}]\,$ \uplambda$6583 emission line, obtained from the modeled spectra. $\mathrm{W_{80}}$ is defined as the width of an emission line that comprises 80 percent of the emitted flux. It is commonly used to identify outflows \citep{zakamska14, Kakkad2020, wylezalek20, ruschel-dutra21}. The [N\,{\sc{ii}}]\,$ \uplambda$6583 line was selected because its emission is observed over the whole GMOS-IFU FoV and other emission lines present similar $\mathrm{W_{80}}$ maps. $\mathrm{W_{80}}$ values larger than 600 km $\mathrm{s^{-1}}$ are usually associated with ionized outflows in quasars \citep[e.g.][]{Kakkad2020}, while values above 500 km $\mathrm{s^{-1}}$ may already trace outflows in lower luminosity AGNs \citep[e.g.][]{wylezalek20, Gatto2024}. For CGCG\,012-070, the $\mathrm{W_{80}}$ map presents values larger than 600 km $\mathrm{s^{-1}}$ mainly in regions south-southwest of the nucleus, with the highest values reaching up to 900 km $\mathrm{s^{-1}}$, which is a signature of gas in gravitationally unbound orbits, indicating the presence of outflow in ionized gas \citep{Sun_2017}.  In all maps, the central cross indicates the position of the nucleus, defined as the position of the peak of the continuum emission.

The bottom panel of Fig.~\ref{fig:large} shows, as an example, the nuclear spectrum (in dashed  black line), the best-fitting model (in red line), and the residuals of the fit are shown in dotted blue line. In general, the observed spectrum of the galaxy is well reproduced by the model, as indicated by the residuals close to the noise level, except for the [Na\,{\sc{i}}]\,$ \uplambda \uplambda$5890,5896 (Na D) line, which often shows a strong absorption contribution from the interstellar medium \citep{Heckman_2000, Roy_2021}. Five inset plots are included to provide a closer view of the emission line regions, highlighting the multi-Gaussian fit, taking into account the stellar population contribution. The narrow and broad components are illustrated as purple dash-dot and green dashed lines, respectively. As can be seen, the emission-line profiles are well reproduced by the Gaussian fits.

Fig.~\ref{fig:maps1_novo} presents the flux distributions and  kinematic maps for the [O\,{\sc{iii}}]\,$\uplambda$5007 and [N\,{\sc{ii}}]\,$\uplambda$6583 emission lines. For each line, the fluxes, centroid velocity and velocity dispersion maps are shown, from left to right. These lines were chosen because they are the strongest ones in each kinematic group. We do not present the kinematic maps for the [O\:{\sc i}] and [N\:{\sc i}] because they are similar to those for the [N\:{\sc ii}]. The velocity fields are shown relative to the systemic velocity of the galaxy and the velocity dispersion maps are corrected for the instrumental broadening. The first two rows correspond to the narrow emission component attributed to gas emission in the disc of the galaxy, while the bottom two rows represent the broad component, associated with outflows. The crosses in all maps mark the position on the nucleus, and the spatial orientation is given in the upper-left panel. Areas shaded in gray indicate spaxels where the emission line is not detected with a signal above 3$\upsigma$ of the noise in regions nearby the emission lines. The remaining emission-line flux maps are shown in Fig.~\ref{fig:maps2_novo}, from left to right, for the [N\,{\sc{i}}]\,$\uplambda$5200, [O\,{\sc{i}}]\,$\uplambda$6300, H$\upalpha$ and [S\,{\sc{ii}}]\,$\uplambda$6731 lines. Top and bottom panels refer to narrow and broad components, respectively.

The flux distributions (Figs.~\ref{fig:maps1_novo} and \ref{fig:maps2_novo})  for the narrow components show extended emission for all emission lines, except for [N\,{\sc{i}}]\,$\uplambda$5200 with detected emission concentrated only at the nucleus. The emission peak for all lines is observed at the nucleus. Among them, the [N\,{\sc{ii}}]\,$\uplambda$6583 line exhibits emission across the entire GMOS FoV, while H$\alpha$ emission is predominantly concentrated within the central 1\,arcsec radius, which corresponds to about 800 pc. Additionally, faint H$\alpha$ emission is detected to the east of the nucleus. [O\,{\sc{iii}}],  [S\,{\sc{ii}}] and [O\,{\sc{i}}] emission is observed mainly in the inner 1 arcsec.  The flux distributions are slightly more elongated along the major axis of the galaxy.  On the other hand, the broad component emission is observed only within the inner 1\,arcsec and exhibits a roughly round flux distribution, except for [N\,{\sc{i}}]\,$\uplambda$5200, which is detected exclusively at the nucleus.

\begin{figure*}
    \centering
    \includegraphics[width=0.99\textwidth, trim=0cm 0.1cm 0cm 0cm, clip]{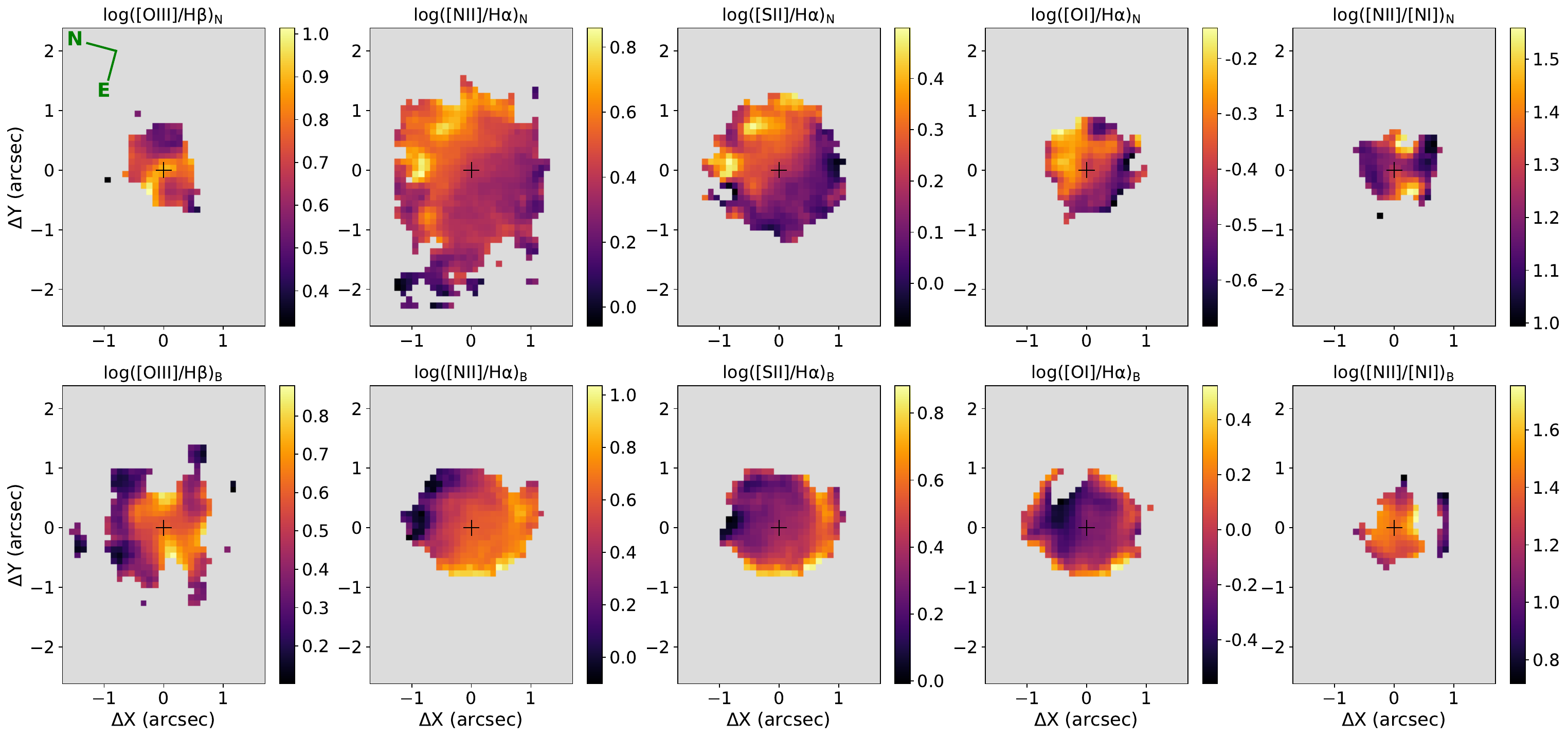} 
    \caption{The panels, from left to right, display maps of [O\,{\sc{iii}}]\,$\uplambda$5007/H$\upbeta$, [N\,{\sc{ii}}]\, $\uplambda$6583/H$\upalpha$, [S\,{\sc{ii}}]\,$\uplambda\uplambda$6716, 6731/H$\upalpha$, [O\,{\sc{i}}]\,$\uplambda$6300/H$\upalpha$, and [N\,{\sc{ii}}]\,$\uplambda$6583/[N\,{\sc{i}}]\,$\uplambda$5200 for the narrow (top) and broad (bottom) components. Gray regions represent masked locations for spaxels where at least one line is not detected with $\mathrm{SNR > 3\upsigma}$.}
    \label{fig:linesratio}
\end{figure*}

The velocity fields of the narrow component in Fig.~\ref{fig:maps1_novo} present a rotation disc pattern, with blueshifts to the east and redshifts to the west of the nucleus, similarly to that observed for the stars (Fig.~\ref{fig:large}). A projected velocity amplitude of $\sim$80  km $\mathrm{s^{-1}}$ is observed for [O\,{\sc{iii}}]\,$\uplambda$5007, while [N\,{\sc{ii}}]\,$\uplambda$6583 shows a velocity amplitude of  $\sim$180  km $\mathrm{s^{-1}}$.  The velocity dispersion maps for the narrow component present a range of values from $\sim$ 70 to 200 km $\mathrm{s^{-1}}$.  
The velocity fields for the broad component of both lines show redshifts of $\sim$80\,km\,s$^{-1}$ to the southeast of the nucleus, while velocities close to the systemic velocity or slightly blueshifted are observed to the northeast of the nucleus.
  Both dispersion velocity maps for the broad component show mild variation, ranging from $\sim$ 370 to 470 km $\mathrm{s^{-1}}$, with the highest values observed mainly to the southeast. 
  
Fig.~\ref{fig:linesratio} shows the observed flux line ratio maps for the narrow (upper panels) and broad (lower panels) components, useful to investigate the gas excitation. The gray regions correspond to masked out spaxels where one or both emission lines of the ratio are not detected with a signal to noise ratio of at least 3. The [N\,{\sc{ii}}]\,$\uplambda$6583/H$\upalpha$, [S\,{\sc{ii}}]\,$\uplambda \uplambda$6716, 6731/H$\upalpha$ and [O\,{\sc{i}}]\,$ \uplambda$6300/H$\upalpha$ line ratios for the narrow component present larger values to the north of the nucleus and lower values to the south, while the [O\,{\sc{iii}}]\,$\uplambda$5007/H$\upbeta$ and [N\,{\sc{ii}}]\,$\uplambda$6583/[N\,{\sc{i}}]\,$\uplambda$5200 ratios present the highest values at the nucleus. All line ratio maps for the broad component show larger values compared to the corresponding narrow component maps, except for [O\, {\sc{iii}}]\,$\uplambda$5007/H$\upbeta$, which exhibits slightly higher values in the narrow component.

\begin{figure}
    \centering
    \includegraphics[width=0.48\textwidth, trim=0cm 0cm 0cm 0cm, clip]{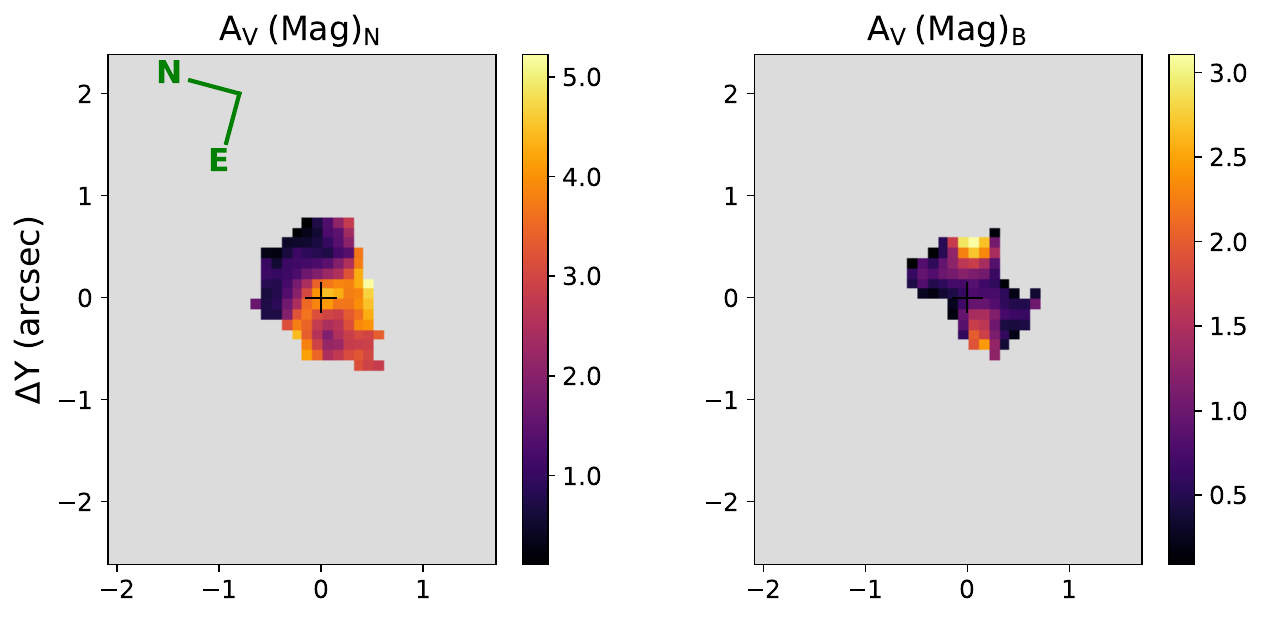} 
    \includegraphics[width=0.48\textwidth, trim=0cm 0cm 0cm 0cm, clip]{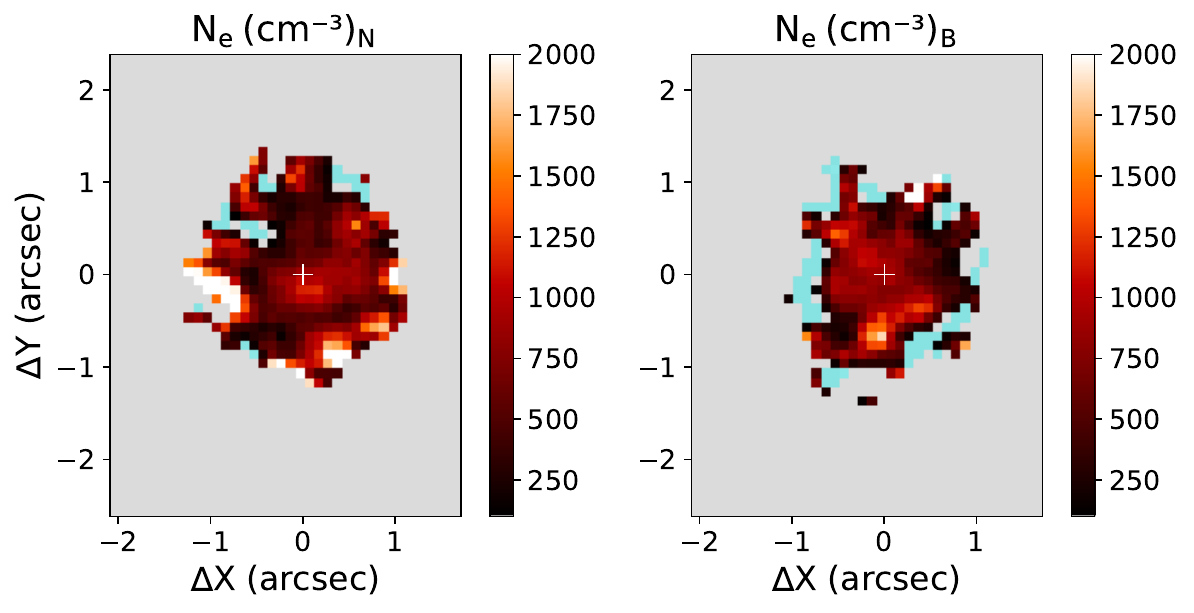} 
    \caption{Top: Visual extinction map for CGCG\,012-070 obtained using H$\upalpha$/H$\upbeta$ ratio for the narrow (left) and broad (right) components. Bottom:   [S\,II]-based electron density maps for narrow (left panel) and broad (right) components. The cyan regions correspond to low-intensity locations traced by $F_{\rm [S\,II]\lambda6717}/F_{\rm [S\,II]\lambda6731} > 0.42$. The gray regions indicate spaxels where one or both emission lines are not detected with a significance of at least 3$\upsigma$ above the continuum noise.}
    \label{fig:Av_maps}
\end{figure}

Following \citet{rogerio_2021}, we estimate the visual gas extinction using the H$\upalpha$/H$\upbeta$ intensity ratio as
\begin{equation}
    A_{\rm V} = 7.22 \log \left[ \frac{(F_{\mathrm{H\alpha}}/F_{\mathrm{H\beta}})_{\mathrm{obs}}}{2.86} \right],
\label{eq:Av}
\end{equation}
where $(F_{\mathrm{H\alpha}}/F_{\mathrm{H\beta}})_{\mathrm{obs}}$ is the observed flux ratio, and we consider the extinction law of \citet{cardelli_1989} and assume an intrinsic flux ratio of $F_{\mathrm{H\alpha}}/F_{\mathrm{H\beta}} = 2.86$, which corresponds to case~B recombination for neutral hydrogen at an electron temperature of $T_{\mathrm{e}} = 10\,000$~K and an electron density of $N_{\rm e} = 100$~cm$^{-3}$ \citep{Osterbrock2006}. The resulting $A_{\rm V}$ maps for the narrow and broad components are shown in the top panels of Fig.~\ref{fig:Av_maps}, covering the inner $\sim$0.7\,arcsec, which corresponds to the region where H$\beta$ emission is detected. 
For the disc component, the $A_V$ presents values ranging from 0.5 to $\sim$4 mag, with the highest values observed to the southeast side of the galaxy, nearly coinciding with the central large-scale dust structure visible in Fig.~\ref{fig:large}. The median visual extinction value for this component is 2.5  mag. The extinction map derived for the outflow component shows lower values, with a median of 0.8\,mag.

The electron density ($N_e$) can be derived from the flux ratio of the [S\,{\sc{ii}}] emission line doublet. To calculate \( N_e \), we used the {\sc PYNEB} Python package \citep{Luridiana2015}, adopting an electron temperature of \( 1.5 \times 10^4 \, \text{K} \), which is characteristic of the narrow-line region in nearby AGNs \citep{revalski18a, revalski21, dors20, riffel21_te, Armah2023}. 
The bottom panels of Fig.~\ref{fig:Av_maps} display the resulting $N_e$ maps obtained for the narrow (left panel) and broad (right panel) components. Cyan areas indicate low-density gas where the [S\,{\sc{ii}}] flux ratio exceeds 1.42, beyond which the ratio is no longer sensitive to density, corresponding to values lower than 100\,cm$^{-3}$ \citep{Osterbrock2006}.  Both density maps for the narrow and broad components show a similar range of values, from $\sim 100$ to 2000\,cm$^{-3}$, with median values of 652\,cm$^{-3}$ and 602\,cm$^{-3}$, respectively \citep[see ][]{Zhang_2024}.

\section{Discussion}
\label{sec: discussion}

As discussed in the previous section, the gas kinematics reveal two distinct components: (i) a rotating disc component, which follows a pattern similar to that observed in the stellar kinematics and is traced by the narrow profiles of the emission lines; (ii) an outflow component, identified through the broad profiles of the emission lines. In this section, we analyse the origin of the gas emission in both components, derive the kinematical properties of the galactic disc in the inner region, and estimate the characteristics of the outflow.

\subsection{Origin of the gas emission}
\label{sec: gasexcitation}

\begin{figure*}
    \centering
    
 \includegraphics[width=0.47\linewidth]{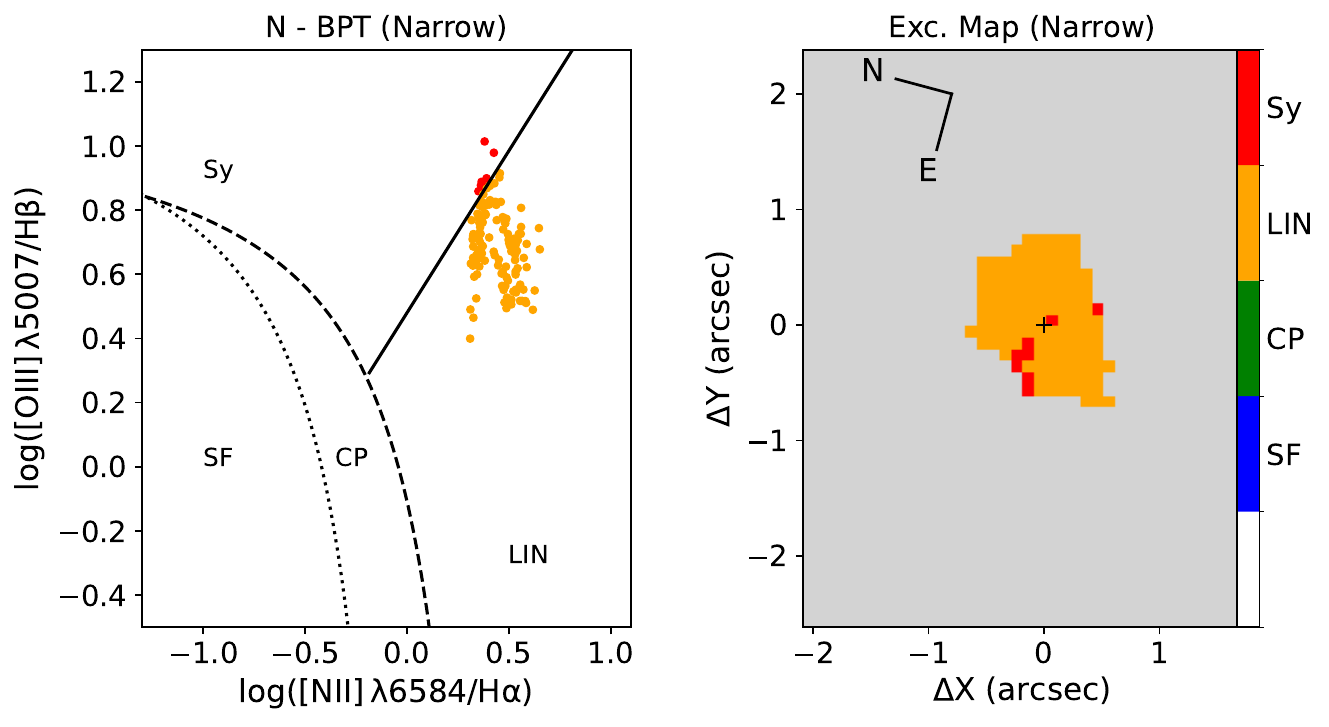}
    \label{fig:bpt_nii_narrow}
    \hfill
 \includegraphics[width=0.47\linewidth]{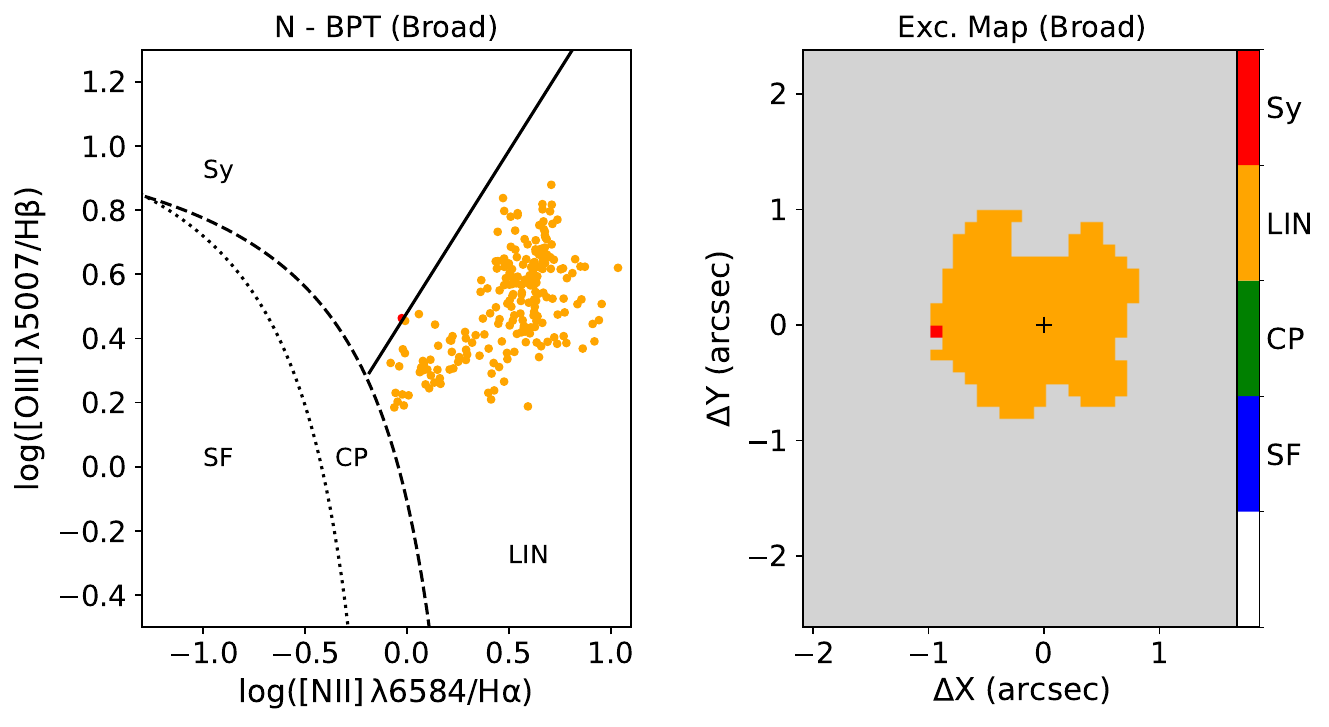}
    \label{fig:bpt_nii_broad}
    
    \vspace{\baselineskip} 
    
 \includegraphics[width=0.47\linewidth]{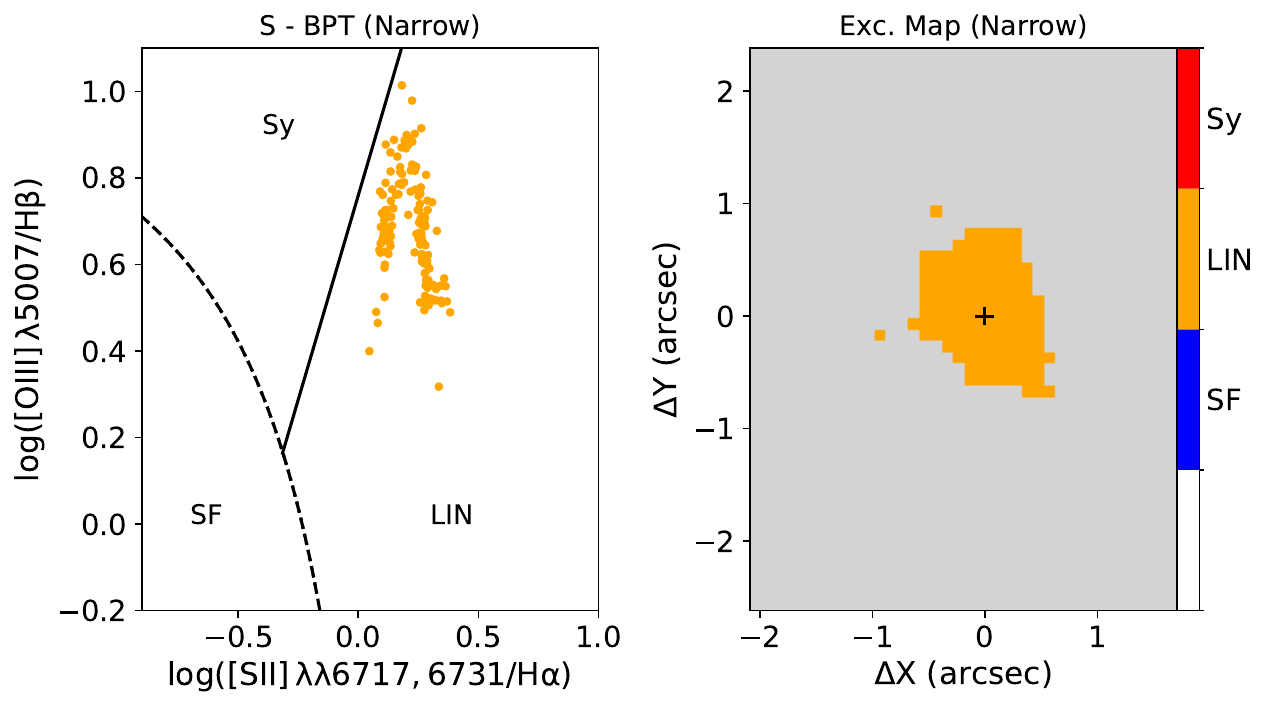}
    \label{fig:bpt_sii_narrow}
    \hfill
 \includegraphics[width=0.47\linewidth]{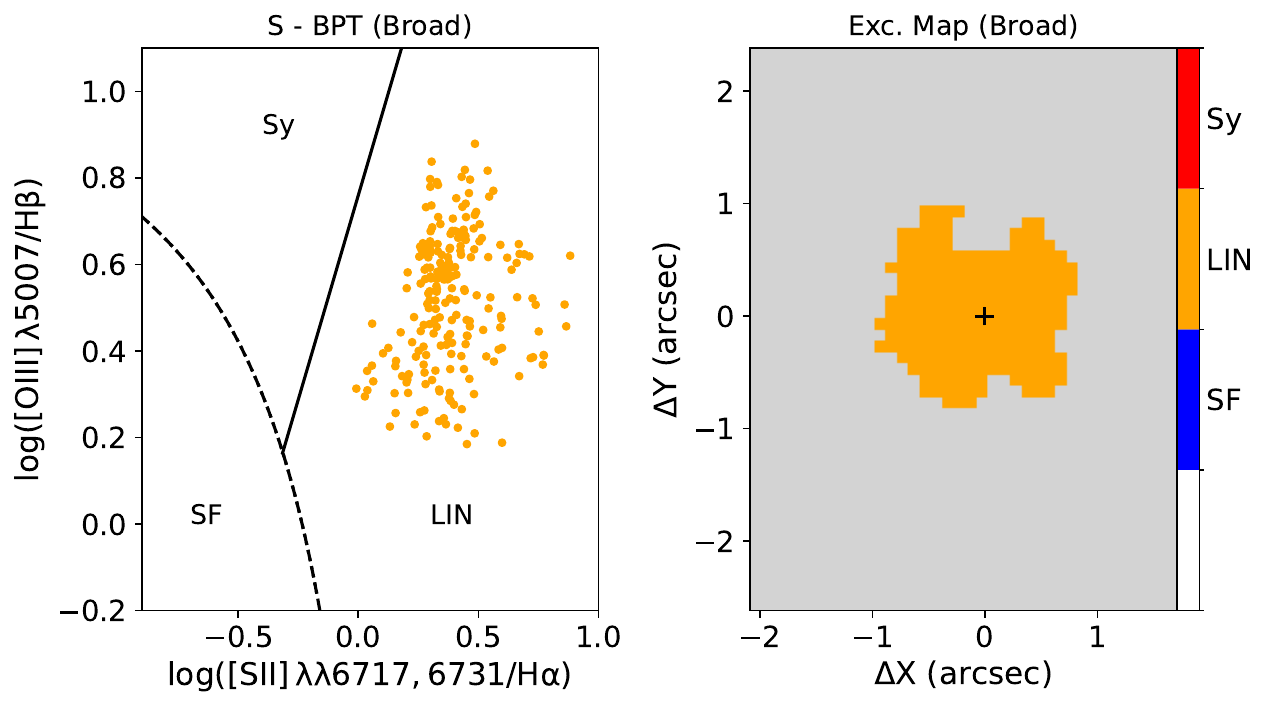}
    \label{fig:bpt_sii_broad}
    
    \vspace{\baselineskip} 
    
 \includegraphics[width=0.47\linewidth]{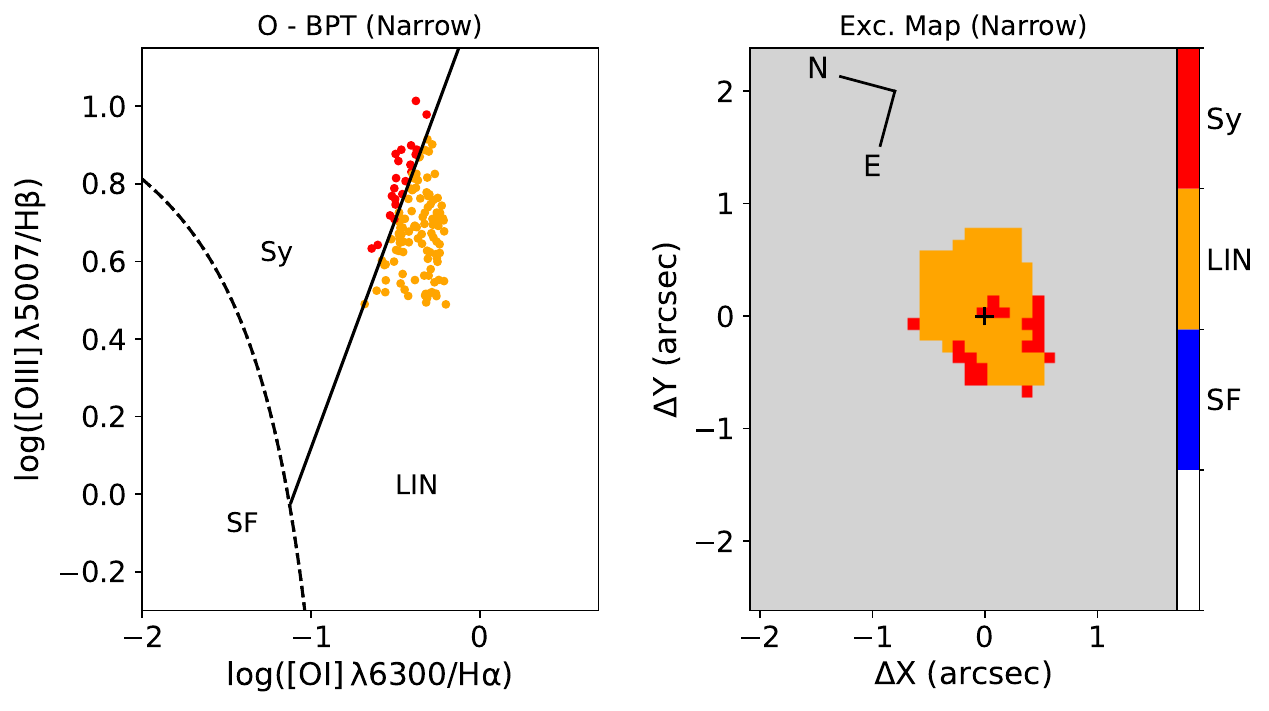}
    \label{fig:bpt_oi_narrow}
    \hfill
 \includegraphics[width=0.47\linewidth]{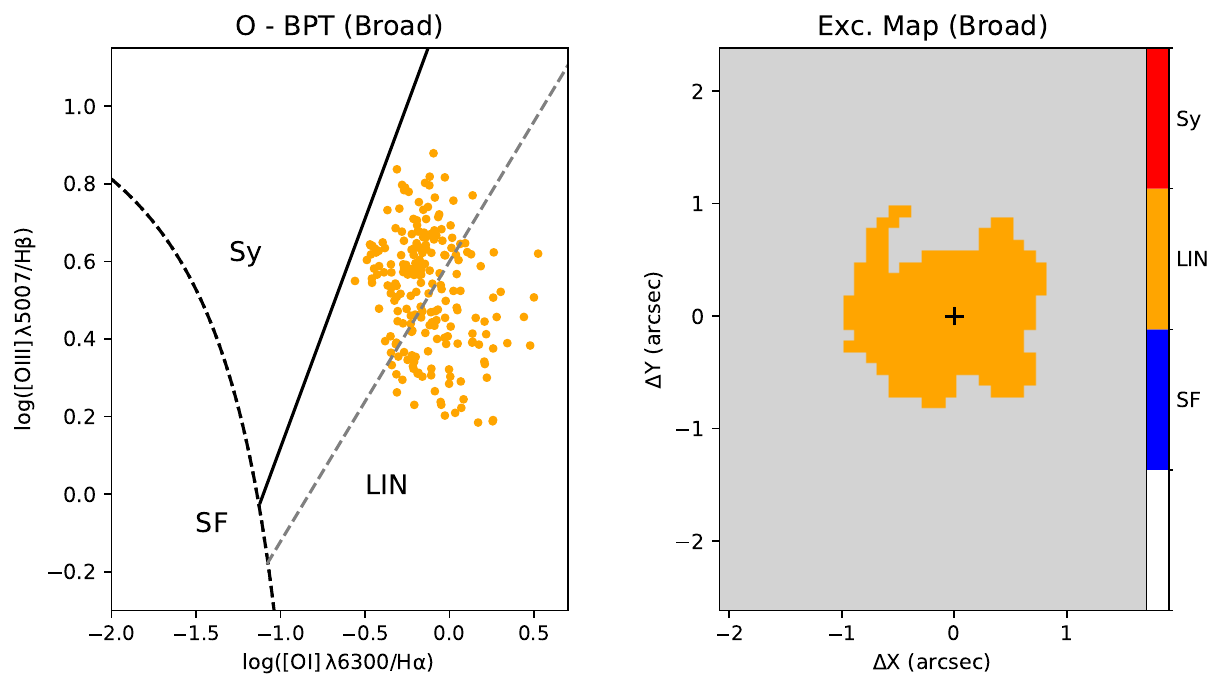}
    \label{fig:bpt_oi_broad}
    
    \caption{Emission-line ratio diagnostic diagrams and excitation maps for CGCG\,012-070. The top row presents the [NII]/H$\alpha$ BPT diagnostic, the middle row displays the [SII]/H$\alpha$ classification, and the bottom row shows the [OI]/H$\alpha$ diagram. In each row, the first two panels correspond to the narrow component, while the last two panels illustrate the broad component. The lines delineate different emission regions: star-forming (SF), composite objects (CP), Seyfert (Sy), and LINER (LIN), following the classifications by \citet{kewley01}, \citet{Kewley_2006}, \citet{kauffmann_2003}, and \citet{Cid_Fernandes_2010}. In the [O\,{\sc i}]-BPT diagram for the broad component, the gray dashed line corresponds to the fiducial model of shock-dominated emission in NGC\,1482 from \citet{Sharp_2010}.}
    \label{fig:BPTs}
\end{figure*}

The use of diagnostic diagrams is a well-established method for classifying the predominant gas ionization mechanisms responsible for the observed emission-line flux ratios. The Baldwin–Phillips–Terlevich \citep[BPT,][]{bpt_1981, veilleux_osterbrock_1987} and WHAN \citep{Cid_Fernandes_2010} diagrams can be used to distinguish gas regions in a galaxy based on their ionization sources. The BPT diagrams for the disc and outflows components in CGCG\,012-070 are shown in Figure~\ref{fig:BPTs}. 
The borderlines in the BPT diagrams distinguish between different emission regions: star-forming (SF), composite objects (CP), Seyfert (Sy) and LINER (LIN) \citep{kewley01, Kewley_2006, kauffmann_2003, Cid_Fernandes_2010}, for narrow and broad components. As observed in the figure, in all three diagrams, the points are concentrated in the AGN region for both the disc and outflow components. These diagrams are restricted to the central region of the galaxy, where all emission lines for each corresponding diagram are detected. Comparing the diagrams for the disc and outflow components, it can be seen that the outflow component presents systematically larger values in the ratios plotted on the x-axis of [N\,{\sc{ii}}]\,$\uplambda$5683/H$\upalpha$, with average values of ${\log \left( \frac{[\ion{N}{ii}]}{\mathrm{H}\upalpha} \right)_N  = 0.41 \pm 0.15}$ and ${ \log \left( \frac{[\ion{N}{ii}]}{\mathrm{H}\upalpha} \right)_B  = 0.52 \pm 0.23}$. In contrast, the disc component of [O\,{\sc{iii}}]\,$\uplambda$5007/H$\upbeta$ has a higher mean than for the outflow one: ${\left\langle \log \left( \frac{[\ion{O}{iii}]}{\mathrm{H}\upbeta} \right)_N \right\rangle = 0.69 \pm 0.12}$; $m{\left\langle \log \left( \frac{[\ion{O}{iii}]}{\mathrm{H}\upbeta} \right)_B \right\rangle = 0.47 \pm 0.17}$.

\begin{figure*}
    \centering
    \includegraphics[width=0.47\linewidth]{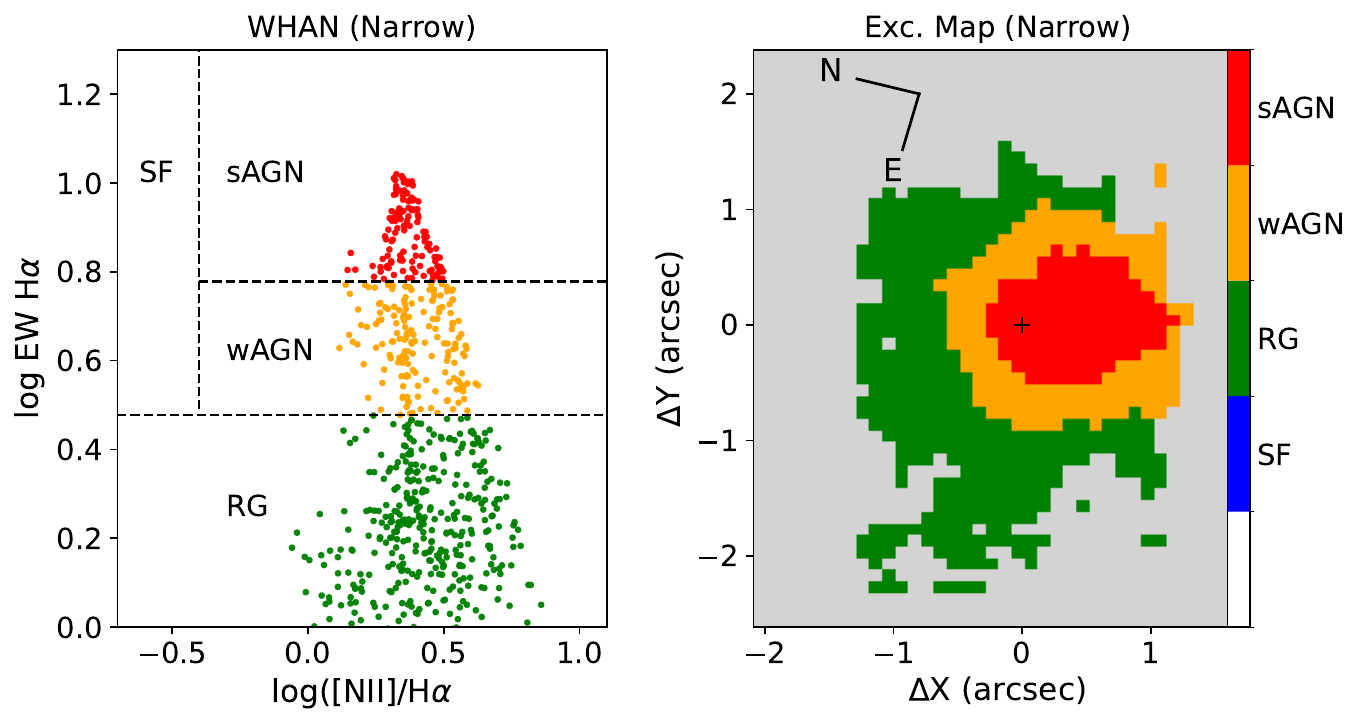}
    \hfill
    \includegraphics[width=0.47\linewidth]{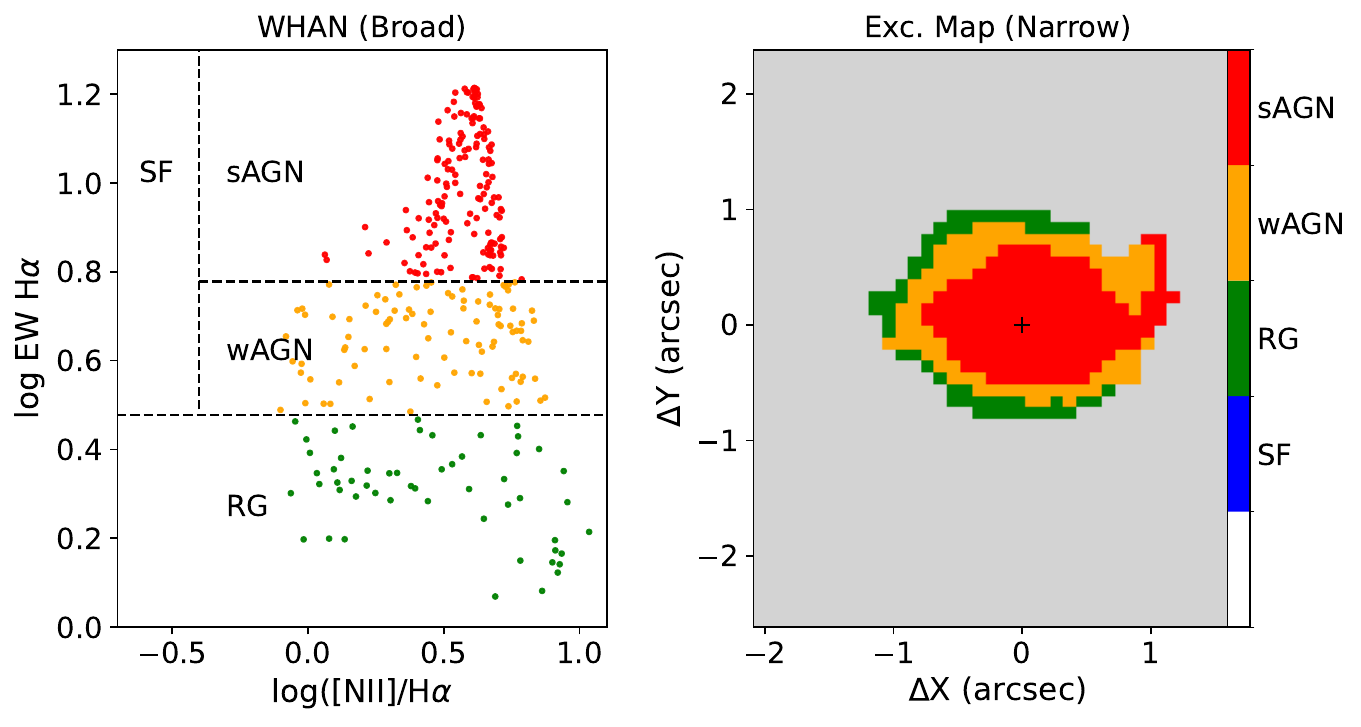}
    \caption{The left panels display the WHAN diagram and the excitation map for the disc component, while the right panels show the corresponding results for the outflow. In the WHAN diagram, the classification lines are based on the criteria from \citet{whan_cid_fernandes_2011}, categorizing regions as star-forming galaxies (SF), weak AGN (wAGN), strong AGN (sAGN), and retired galaxies (RG).}
    \label{fig:Whan}
\end{figure*}

The classification of LINERs in the BPT diagram may include contributions from non-AGN sources such as hot low-mass evolved stars responsible for the ionization observed in retired galaxies (RG) that have stopped forming stars \citep[HOLMES,][]{Cid_Fernandes_2010,whan_cid_fernandes_2011,agostino21}. To unravel this mixed emission, the WHAN diagram is commonly employed, taking into account the H$\upalpha$ equivalent width (EW$\mathrm{_{H\upalpha}}$) versus [N\,{\sc{ii}}]\,/H$\upalpha$ ratio. The WHAN diagrams and excitation maps are shown in Fig.~\ref{fig:Whan}, for the disc (first two panels) and outflow (last two panels) components. As the WHAN diagram is not limited to the region where H$\beta$ is detected, as is the case for the BPT diagrams, the excitation maps cover a larger region of the galaxy. The lines in the WHAN diagram follow the definitions provided by \citet{whan_cid_fernandes_2011}, and the classifications used are: star-forming galaxies (SF), weak AGN (wAGN), strong AGN (sAGN), and retired galaxies (RG). The WHAN diagrams confirm the presence of an AGN in the centre of CGCG\,012-070, with points located in the sAGN region, transitioning to wAGN and later to RG in the outermost regions, a behavior expected due to the dilution of the AGN radiation field with distance from it. This is consistent with the classification based on integrated optical spectra, where the nuclear activity of CGCG\,012-070 is identified as a Sy 2 nucleus \citep{veron06,Thomas13}.

As mentioned above, the values of [N\,{\sc{ii}}]/H$\upalpha$, [S\,{\sc{ii}}]\,$\uplambda\uplambda$6716,6731/H$\upalpha$, and [O\,{\sc{i}}]\,$\uplambda$6300/H$\upalpha$ for the outflow component are higher than those for the disc component. The median logarithmic values for these ratios are 0.39, 0.18, and -0.40 for the disc component, whereas for the outflow component, they are 0.58, 0.38, and -0.13, respectively. This indicates that the low-ionization lines in the outflow have an additional excitation mechanism besides photoionization by the central source. The most likely additional mechanism is fast shocks, which are more efficient at producing emission from low-ionization lines such as [S\,{\sc ii}], [N\,{\sc ii}], and [O\,{\sc i}] compared to higher-ionization lines like [O\,{\sc iii}] \citep[e.g.][]{allen08}. We can also compare the gas ionization degree of the outflow and disc gas using the [O\,{\sc{iii}}]\,$\uplambda$5007/[N\,{\sc{ii}}]\,$\uplambda$6583 and [O\,{\sc{iii}}]\,$\uplambda$5007/[S\,{\sc{ii}}]\,$\uplambda\uplambda$6716,6731 line ratios. By integrating the spaxels detected in both lines and correcting them for extinction, we find the following values for narrow e broad components of the flux ratios: [O\,{\sc{iii}}]/[N\,{\sc{ii}}]$_N$\,=\,0.58\,$\pm\,0.07$; [O\,{\sc{iii}}]/[N\,{\sc{ii}}]$_B$\,=\,0.48\,$\,\pm\,0.06$; [O\,{\sc{iii}}]/[S\,{\sc{ii}}]$_N$\,=\,1.1\,$\pm\, 0.13$ and [O\,{\sc{iii}}]/[S\,{\sc{ii}}]$_B$\,=\,0.77\,$\pm\,0.09$. Therefore, the outflow gas presents a slightly lower ionization degree than the gas at the disc, which further supports the contribution of shocks to the gas ionization, as shocks are more efficient in producing low-ionization lines.

Since CGCG\,012-070 exhibits only mild radio emission \citep{Lofthouse2018, Becker1995}, the most probable cause of the shocks are the outflows, as observed in other galaxies \citep[e.g.][]{Ho_2014,DAgostino19a,DAgostino19b,Riffel2023,Riffel2024}. This is further corroborated by the [O\,{\sc{i}}]--BPT diagram for the outflow component (Fig.~\ref{fig:BPTs}), where most points are located in the locus of shock-excited emission for NGC\,1482, considered a prototype of shock-dominated sources \citep{Veilleux02,Sharp_2010}. Thus, we conclude that the gas emission for the disc component is produced by AGN photoionization, while for the outflow component an additional contribution from shocks induced by outflows is necessary to reproduce the observed line ratios in CGCG\,012-070.

\subsection{The disc component}

\label{sec: disccomponent}

The stellar (Fig.~\ref{fig:large}) and narrow-component gas (Fig.~\ref{fig:maps1_novo}) velocity fields clearly display a rotation pattern, with the north-west side approaching and the south-east side receding. In order to estimate the geometric parameters of the disc and characterize the stellar and gas motions, we fit the stellar and gas velocity fields using an analytical model, assuming circular orbits confined to the plane of the galaxy and moving under the influence of a central gravitational potential \citep{bertola_1991}. In this model, the rotational velocity field is defined as

\begin{equation}
\scalebox{0.98}{$
    V(R, \psi) = V_{\mathrm{sys}} + 
    \frac{A R \cos(\psi - \psi_0) \sin \theta \cos^p \theta}
    {\left\{ R^2 \left[ \sin^2 (\psi - \psi_0) + \cos^2 \theta \cos^2 (\psi - \psi_0) \right] + c_0 \cos^2 \theta \right\}^{p/2}}
$}
\label{eq:Bertola}
\end{equation}

\noindent where $V_{\rm sys}$ is the heliocentric systemic velocity of the galaxy, $A$ is the velocity amplitude, $R$ is the distance of each spaxel to the centre of rotation, $\psi$ is the position angle (PA) of each spaxel, $\psi_0$ is the PA of the line of the nodes,  $\theta$ is the inclination of the galactic disc relative to the plane of the sky ($\theta = 0$ for a face-on disc), and $c_0$ is a concentration parameter, defined as the radius where the velocity reaches 70\% of the velocity amplitude. The $p$ parameter represents the slope of the rotation curve, varying from 1.0 for an asymptotically flat rotation curve to 1.5 for a system with finite mass, as for a Plummer potential \citep{Plummer_1911}. We applied the Levenberg-Marquardt algorithm for the fitting of nonlinear least squares using the {\sc levmarlsqfitter} package from the {\sc astropy} Python package \citep{astropy_Robitaille_2013}.

\begin{figure}
    \centering
    \includegraphics[width=0.45\textwidth]{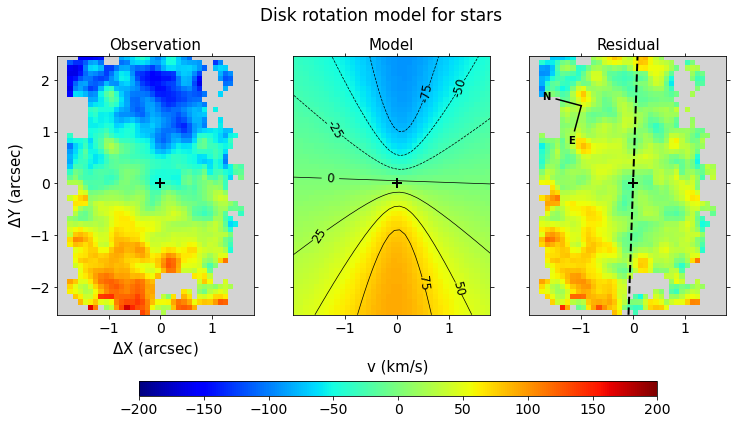}
    \vspace{0.3cm} 

    \includegraphics[width=0.45\textwidth]{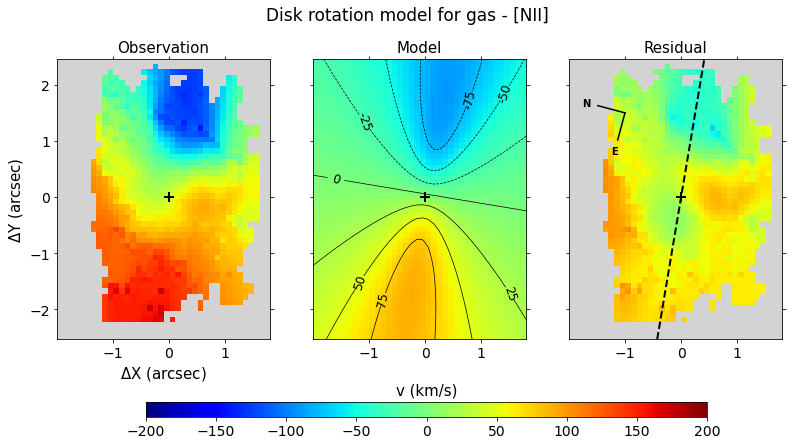}
    
    \caption{Observed velocity maps (left), disc rotation models (centre), and residuals (right) for the stars (top) and the [N,{\sc ii}],$\uplambda$6583 narrow component (bottom). The black dashed line in the residual maps indicates the line of nodes, as determined from the best-fit position angle $\psi_{0}$.}
    \label{fig:rotation}
\end{figure}

In the left panels of Fig.~\ref{fig:rotation}, we present the observed velocity field of the stars (top) and for the narrow component of [N\,{\sc{ii}}]\,$\uplambda$6583 (bottom). The corresponding rotation disc models are shown in the middle panels, while the residual maps -- obtained by subtracting the model from the observed velocities -- are in the right panels. The best-fit parameters are listed in Table~\ref{tab:modelo_rotacao}. For the gas, $\theta$, $V_{\rm sys}$ and $c_0$ were fixed to the values derived from the stellar component.  The best-fit parameters for the stars and gas are similar and in agreement with values obtained for the large scale disc \citep[e.g.][]{Skrutskie_2006_2MASS}.

\begin{table}
\centering
\caption{Best-fitting parameters of the rotation disc models for stars and ionized gas. }
\label{tab:modelo_rotacao}
\begin{tabular}{lcc}
\hline
Parameter & Stars & Ionized Gas \\
\hline
$A$ (km\,s$^{-1}$) & $150 \pm 31$ & $243 \pm 31$ \\
$c_0$ (arcsec)     & $1.0 \pm 0.6$   & fixed \\
$p$                & $1.10 \pm 0.06$ & $1.29 \pm 0.04$ \\
$\psi_0$ (degrees) & $103 \pm 4$  & $95 \pm 6$ \\
$\theta$ (degrees) & $67.4 \pm 8.5$  & fixed \\
$V_\mathrm{sys}$ (km\,s$^{-1}$) & $14\,480 \pm 37$ & fixed \\
\hline
\end{tabular}
\end{table}

The residual map for the stars shows low values ($\lesssim20$ km\,s$^{-1}$) at most locations, except for a region approximately 2 arcsec east of the nucleus, where the residuals reach redshifted velocities of $\sim100$ km\,s$^{-1}$ in an arc-shaped structure. Similar values are observed in the gas residual map in the same region and some blueshifted residuals ($\sim50$ km\,s$^{-1}$) west of the nucleus, co-spatial with a region of enhanced velocity dispersion in the narrow [N\,{\sc ii}] component. A plausible explanation for these residuals is a recent capture of a small satellite galaxy, as commonly occurs in nearby AGN hosts \citep[e.g.][]{davis11,raimundo17,raimundo21,raimundo22,Riffel2024, Rembold_2024}. In addition, the residual map for the gas shows redshifts to the south of the nucleus, approximately along the region where the highest $W_{\rm 80}$ values are observed (Fig.~\ref{fig:large}). 
 These residuals are possibly associated with the interaction of the outflow with the ISM on the near side of the galaxy disc.

\subsection{Outflow properties}
\label{sec: outflows}

The large-scale image of CGCG\,012-070 suggests that the southwest side is slightly more obscured by dust, indicating that it represents the near side of the disc, while the northeast side corresponds to the far side. The velocity fields of the broad component exhibit redshifts toward the south of the nucleus and some blueshifts toward the north (Fig.~\ref{fig:maps1_novo}). Additionally, the $W_{\rm 80}$ map (Fig.~\ref{fig:large}) shows higher values to the south of the nucleus, which we interpret as resulting from the interaction between the outflow and the gas in the disc. Based on this, the broad component observed in the emission lines is consistent with a biconical outflow roughly aligned in the north-south direction, with redshifts seen on the southern side behind the disc plane and blueshifts in the northern side in front of the disc plane. The small velocity gradient indicates that the outflow is approximately in the plane of the sky, consistent with the galaxy disc orientation.
In this section, we estimate the ionized gas outflow properties and discuss their impact on the host galaxy.

We estimate the mass of the ionized gas outflow ($M_{\rm out}$) by \citep{Carniani_2015}:
\small{
\begin{equation}
M_{\rm out} = 0.8 \times 10^8\,{\rm M}_\odot\, 
\frac{1}{10^{[\rm O/H] - [\rm O/H]_\odot}}\, 
\frac{L_{\text{[O\,\sc{iii}}],\, \rm broad}}{10^{44}\,{\rm erg\,s}^{-1}}\, 
\frac{ \Tilde{n}_e }{500\,{\rm cm}^{-3}} ,
\label{eq:massa}
\end{equation}} \normalsize

\noindent where $\Tilde{n}_e$ is the median electron density for the outflow estimated from the flux ratios of the [S\,{\sc{ii}}] broad components and $L_{\text{[O\,{\sc{iii}}}, \, \rm broad]}$ is the integrated luminosity of the broad component of [O\,{\sc{iii}}]\,$\uplambda$5007, corrected by extinction using the median value of $A_{\rm V,\,broad}$ for the broad component and adopting the extinction law from \citet{cardelli_1989}. The oxygen abundance [O/H] can be estimated from the log([N\,{\sc{ii}}]\,$\uplambda$6583/H$\upalpha$) calibration for nearby LINER galaxies derived by \citet{Oliveira22}. Using the fluxes for the broad component and a solar oxygen abundance of 12 + [O/H]$_{\odot}$=8.69 \citep{Allende_Prieto_2001}, we obtain [O/H]-[O/H]$_{\odot}$=0.30 for the outflow in CGCG\,012-070.

Then, we follow \citet{Riffel_2023_AGNIFS}, \citet{Bruno_Oliveira_2021} and  \citet{Gatto2024} to estimate the outflow properties. The velocity of the outflow ($V_{\rm out}$) is defined as 
\begin{equation}
V_{\rm out} = \frac{\langle (|v_{\rm broad}| + 2\sigma_{\rm broad})\, F_{\rm broad} \rangle}{\langle F_{\rm broad} \rangle} ,
\label{eq:v_out}
\end{equation}
\noindent where $F_{\rm broad}$, $v_{\rm broad}$ and $\sigma_{\rm broad}$ are, respectively, the flux, centroid velocity and velocity dispersion of the [O\,{\sc{iii}}]\,$\uplambda$5007 broad component in each spaxel. 
Similarly, the radius of the outflow ($R_{\rm out}$) is derived from a flux-weighted average of the spatial extent of the broad [O\,{\sc iii}] component:
\begin{equation}
R_{\rm out} = \frac{\langle R_{\rm broad}\, F_{\rm broad} \rangle}
{\langle F_{\rm broad} \rangle}.
\end{equation}
\noindent Here, $R_{\rm broad}$ is the distance of each spaxel from the nucleus. Then, the mass-outflow rate is estimated by 
\begin{equation}
\dot{M}_{\rm out} = 
\frac{M_{\rm gas}\, V_{\rm out}}{R_{\rm out}}.
\end{equation}

The kinetic power of the outflow is computed as 
\begin{equation}
\dot{E}_{\rm out} = 
\frac{1}{2} \dot{M}_{\rm out} \, V_{\rm out}^2.
\end{equation}

Using the equations above, we obtain the following values: $M_{\rm out}~=~(4.6\pm0.51)\times 10^{4}M_{\odot}$; $V_{\rm out}~=~(1330 \pm 13)\rm\,km\,s^{-1}$; $R_{\rm out} = (0.94\pm0.36)$ kpc; $\dot{M}_{\rm out} = (0.067 \pm 0.026)\,M_{\odot} \,yr^{-1}$; and $\dot{E}_{\rm out} = (3.7 \pm 1.5) \times 10^{40}\, \rm erg\,s^{-1}$. The uncertainties reported for the outflow properties represent only measurement errors and do not account for systematic uncertainties related to the outflow geometry or assumptions about the outflow density, which could be significant as extensively discussed in the literature \citep[e.g.][]{Davies2020,revalski22,Riffel_2023_AGNIFS,Revalski2025}.

As CGCG\,012-070 presents only a weak radio source \citep[$P_{1.4} \approx 2.4 \times 10^{22}$ W Hz$^{-1}$;][]{Lofthouse2018, Becker1995}, the ionized outflows are likely driven by the radiation pressure exerted by the accretion disc surrounding the SMBH.
To further investigate the physical processes responsible for driving the observed outflows, we calculate the ratio between the momentum flux of the outflow ($\dot{P}_{\rm out}$) and photon momentum flux ($\dot{P}_{\rm AGN}$) via 
\begin{equation}
        \frac{\dot{P}_{\rm out}}{\dot{P}_{\rm AGN}} = \frac{\dot{M} \times V_{\rm out}}{L_{\rm bol}/c},
\label{eq:ratio_P}
\end{equation}
 where $c$ is the light speed, $L_{\mathrm{bol}}$ is the AGN bolometric luminosity, estimated as $L_{\rm bol}/[10^{40}\,{\rm erg\,s}^{-1}] = 
112 (L_{\text{[O\,\sc{iii}}]}/[10^{40}\,{\rm erg\,s}^{-1}])^{1.2}$ \citep{Trump_2015} and $L_{\text{[O\,\sc{iii}]}}$ is the total [O\,{\sc{iii}}]\,$\uplambda$5007 luminosity, corrected by extinction. The ratio $\dot{P}_{\rm out}/\dot{P}_{\rm AGN} \approx 0.3$ for CGCG\,012-070 provides further support that the observed ionized outflows are driven by radiation pressure from the AGN. This value is consistent with radiation-pressure driven winds occurring in low-density environments or with shocked AGN winds, with $\dot{P}_{\rm out} \lesssim 1\,\dot{P}_{\rm AGN}$ \citep{Faucher-Giguere_2012}.

To investigate the impact of the ionized outflow on the host galaxy, we estimate its kinetic coupling efficiency $\epsilon_f = \dot{E}_{\rm out}\,/\,L_{\rm bol} $. The resulting AGN bolometric luminosity and outflow coupling efficiency are $L_{\rm bol}\approx 5.4 \times 10^{43}\, \rm erg\,s^{-1}$ and $\epsilon_f = 0.068 \%$.  

\begin{figure}
    \centering
    \includegraphics[width=0.48\textwidth, trim=0.5cm 0.1cm 1cm 2.4cm, clip]{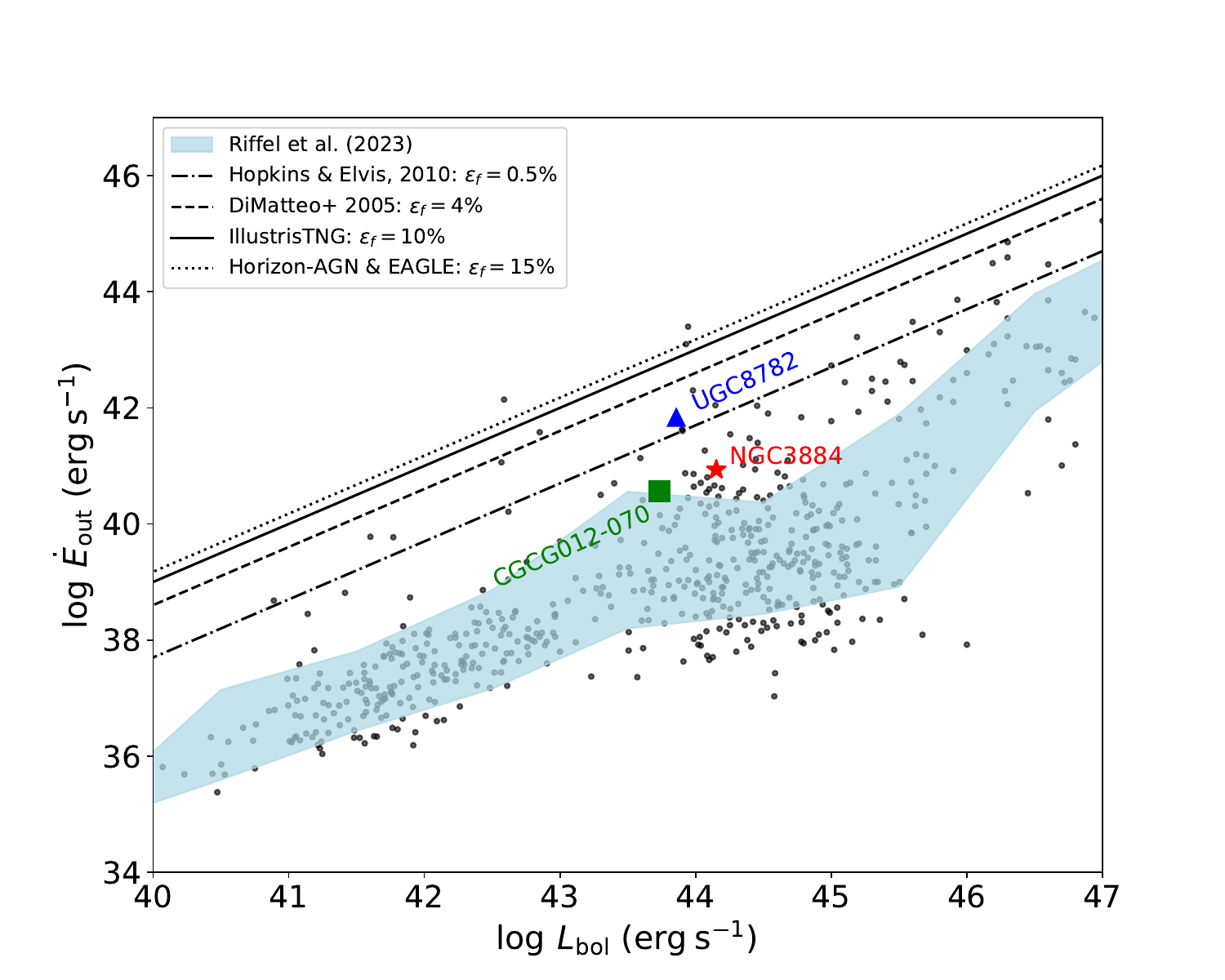}
    \caption{Plot of the kinetic power of ionized outflows as a function of the AGN bolometric luminosity, based on the data from \citet{Riffel_2021_AGNIFS}, represented by black dots. The light blue shaded area corresponds to the mean kinetic power of ionized outflows along with the $\pm 1\sigma$ dispersion, computed within bolometric luminosity bins of 0.5 dex. The lines indicate the minimum coupling efficiencies required for outflows to effectively suppress star formation, as predicted by various simulations \citep{dimatteo05, hopkins_elvis10, dubois_horizon_14, Harrison_2018,Xu_2022}. The blue triangle marks the position of UGC 8782 \citep{Riffel2023}, the red star corresponds to NGC 3884 \citep{Riffel2024}, and the green square indicates CGCG\,012-070.}
    \label{fig:Ekin}
\end{figure}

Theoretical models predict that radiatively-driven AGN outflows must achieve coupling efficiencies of at least 0.5--15\% of $L_{\rm bol}$ for AGN feedback to effectively suppress star formation in their host galaxies \citep{dimatteo05, hopkins_elvis10, dubois_horizon_14, Harrison_2018, Xu_2022}. The value obtained above for the kinetic coupling efficiency of the ionized gas outflows in CGCG\,012-070 falls below these thresholds. 
Fig.~\ref{fig:Ekin} shows the relation between the kinetic power of ionized outflows and AGN bolometric luminosity, based on data compiled by \citet{Riffel_2021_AGNIFS}, which includes measurements for over 600 sources spanning seven orders of magnitude in luminosity. The blue shaded region represents the range between the mean kinetic power of the outflows minus one standard deviation and the mean plus one standard deviation, calculated within 0.5 dex bins of $L_{\rm bol}$.  Besides CGCG\,012-070, results for UGC\,8782 \citep{Riffel2023} and for NGC\,3884 \citep{Riffel2024} are also included in the plot.

AGN-driven outflows are observed across multiple gas phases, with cold molecular outflows generally displaying mass rates and kinetic powers higher than those of their ionized counterparts in galaxies with AGN bolometric luminosity lower than 10$^{45} \rm{erg\,s^{-1}}$ \citep{Fiore2017, Vayner_2021, Harrison_2018, Harrison2024, Riffel_2023_AGNIFS, Fluetsch_2018}. For CGCG\,012-070, the ionized outflows exhibit a coupling efficiency lower than the threshold predicted by the simulations to effectively suppress star formation in the host galaxy. Thus, to further investigate the role of AGN feedback in CGCG\,012-070, other gas phases must be taken into account. 
Moreover, the kinetic energy injected into the interstellar medium by the outflows accounts for only $\sim$20\% of their total coupling energy \citep{richings18b-shocks}. Thus, accounting for the total coupling efficiency of the multi-phase outflows could significantly enhance the overall effectiveness of the outflow in suppressing star formation. Finally, we have shown that CGCG\,012-070 hosts a low-luminosity AGN, where simulations of low-power winds suggest that such winds can effectively quench star formation in the galaxy if they are sustained for longer than approximately 1 Myr \citep{almeida23}.

\section{Conclusions}
\label{sec: conclusions}

We employed optical IFU observations to map the gas emission, extinction, excitation, and kinematics, as well as the stellar kinematics, within the central $3.6 \times 5.1\, \mathrm{kpc}^2$ region of the galaxy CGCG\,012-070.  Our main conclusions are summarized as follows:

\begin{itemize}
    \item The stellar velocity field presents a well-defined rotation pattern, with a projected amplitude of around 200 km\,s$^{-1}$, presenting blueshifts on the northwest side of the galaxy and redshifts on the southeast. The observed velocities are well-represented by a rotating disc model, with the major axis oriented approximately along the position angle $\Psi_0 \approx 103^\circ$, consistent with the large-scale orientation of the disc. 

    \item The gas exhibits two distinct kinematic components, observed as narrow and broad components in the emission-line profiles. The narrow component originates from gas in the disc, following motions similar to those of the stars. The broad component is attributed to an ionized outflow, observed within the inner $1.5\, \mathrm{kpc}$ radius.
 
    \item The gas emission in the disc component is primarily produced by photoionization from the central AGN, while the outflow shows additional emission from shocked gas, as indicated by the emission-line ratio diagnostic diagrams.

    \item The median visual extinction, derived from the Balmer decrement, is $2.5 \, \mathrm{mag}$ for the disc component and $0.8 \, \mathrm{mag}$ for the outflowing gas. The median electron densities, estimated from the [S\,{\sc ii}] doublet, are $652 \, \mathrm{cm}^{-3}$ and $602 \, \mathrm{cm}^{-3}$ for the disc and outflow, respectively.

    \item The ionized outflows are radiatively driven, as suggested by the comparison between the momentum flux of the outflow and the photon momentum flux, with the latter being approximately three times larger than the former. This conclusion is further supported by the absence of a radio jet in the galaxy, based on the available radio data.
    
    \item We estimate a mass outflow rate of $\approx 6.7 \times 10^{-2} \, \mathrm{M_\odot \, yr^{-1}}$ and a kinetic power of $\approx 3.7 \times 10^{40} \, \mathrm{erg \, s^{-1}}$. This corresponds to a kinetic coupling efficiency of the outflows with the ISM of $\sim 0.07$\%. This value is below the outflow coupling efficiencies required by cosmological simulations for AGN feedback to effectively suppress star formation in their host galaxies. However, it may still be sufficient to quench star formation in the central region if sustained over timescales of $\sim 1 \, \mathrm{Myr}$ at the current power, indicating the action of a maintenance mode AGN feedback in this galaxy.
\end{itemize}

In summary, the three galaxies--UGC\,8782 \citep{Riffel2023}, NGC\,3884 \citep{Riffel2024}, and CGCG\,012-070 (this work)--selected for exhibiting high H$_2$/PAH ratios in the mid-infrared and elevated [O\,{\sc i}] velocity dispersions, show clear evidence of outflow components, with shocks playing an important role in the origin of their emission. The ionized gas outflows in these three galaxies are more powerful than those typically observed in AGNs of similar luminosities. Furthermore, it is necessary to explore other gas phases to fully assess the true impact of the outflows on star formation in their host galaxies.

\section*{Acknowledgements}
The authors thank an anonymous referee for the valuable comments and suggestions that helped improve this paper. 
LRV thanks partial financial support from Coordena\c{c}\~ao de Aperfei\c{c}oamento de Pessoal de N\'{\i}vel Superior (CAPES; Finance code 001 \& Proj. 88887.894973/2023-00). RAR acknowledges the support from Conselho Nacional de Desenvolvimento Cient\'ifico e Tecnol\'ogico (CNPq; Proj. 303450/2022-3, 403398/2023-1, \& 441722/2023-7), Funda\c c\~ao de Amparo \`a pesquisa do Estado do Rio Grande do Sul (FAPERGS; Proj. 21/2551-0002018-0), and CAPES  (Proj. 88887.894973/2023-00). RR acknowledges support from  CNPq (Proj. 445231/2024-6,311223/2020-6, 404238/2021-1, and 310413/2025-7),  FAPERGS  (Proj. 19/1750-2 and 24/2551-0001282-6) and   CAPES (Proj. 88881.109987/2025-01). ChatGPT (GPT-4.5) and DeepSeek (V3.1) were used to help debug codes and improve sentence wording. 
Based on observations obtained at the Gemini Observatory, which is operated by the Association of Universities for Research in Astronomy, Inc., under a cooperative agreement with the NSF on behalf of the Gemini partnership: the National Science Foundation (United States), National Research Council (Canada), CONICYT (Chile), Ministerio de Ciencia, Tecnolog\'{i}a e Innovaci\'{o}n Productiva (Argentina), Minist\'{e}rio da Ci\^{e}ncia, Tecnologia e Inova\c{c}\~{a}o (Brazil), and Korea Astronomy and Space Science Institute (Republic of Korea).  This research has made use of NASA's Astrophysics Data System Bibliographic Services. This research has made use of the NASA/IPAC Extragalactic Database (NED), which is operated by the Jet Propulsion Laboratory, California Institute of Technology, under contract with the National Aeronautics and Space Administration.

\section*{Data Availability}

The data used in this paper are available in the Gemini Observatory Science Archive (\url{https://archive.gemini.edu/}) under the program code \texttt{GS-2021B-Q-220}.

\bibliographystyle{mnras}
\bibliography{refs}

\bsp	
\label{lastpage}
\end{document}